\documentclass[preprint]{aastex}

\begin{document}

\title{Nuclear Stellar Populations in the Infrared
Space Observatory Atlas of Bright Spiral Galaxies}

\author{George J. Bendo,\altaffilmark{1,2,3}
Robert D. Joseph,\altaffilmark{2,4}}

\altaffiltext{1}{Steward Observatory, 933 North Cherry Avenue, Tucson, AZ
85721, USA; gbendo@as.arizona.edu}
\altaffiltext{2}{Visiting Astronomer at the Infrared Telescope Facility,
which is operated by the University of Hawaii under contract from the
National Aeronautics and Space Administration.}
\altaffiltext{3}{Visiting Astronomer at the UH 2.2 m Telescope at Mauna Kea
Observatory, Institute for Astronomy, University of Hawaii.}
\altaffiltext{4}{University of Hawaii, Institute for Astronomy,
2680 Woodlawn Drive, Honolulu, HI 96822, USA; joseph@ifa.hawaii.edu}

\shorttitle{Nuclear Stellar Populations}
\shortauthors{Bendo \& Joseph}

\begin{abstract}
To understand the nuclear stellar populations and star formation
histories of the nuclei of spiral galaxies, we have obtained K-band
nuclear spectra for 41 galaxies and H-band spectra for 20 galaxies in
the ISO Atlas of Bright Spiral Galaxies.  In the vast majority of the
subsample (80\%), the near-infrared spectra suggest that evolved red
stars completely dominate the nuclear stellar populations and that hot
young stars are virtually non-existent.  The signatures of recent star
formation activity are only found in 20\% of the subsample, even though
older red stars still dominate the stellar populations in these
galaxies.  Given the dominance of evolved stars in most galaxy nuclei
and the nature of the emission lines in the galaxies where they were
detected, we suggest that nuclear star formation proceeds in the form
of instantaneous bursts.  The stars produced by these bursts comprise
only $\sim$2 \% of the total nuclear stellar mass in these galaxies, but we
demonstrate how the nuclear stellar populations of normal spiral
galaxies can be built up through a series of these bursts.  The bursts
were detected only in Sbc galaxies and later, and both bars and
interactions appeared to be sufficient but not necessary triggers for
the nuclear star formation activity.  The vast majority of galaxies
with nuclear star formation were classified as HII galaxies.  With one
exception, LINERs and transition objects were dominated by older red
stars, which suggested that star formation was not responsible for
generating these galaxies' optical line emission.
\end{abstract}


\section{Introduction \label{s_intro}}

The nuclear stellar populations of galaxies can reveal important
information on the history of nuclear star formation, which could then
reveal the mechanisms behind triggering star formation.  Understanding
nuclear stellar populations is therefore important to understanding
the evolution of galaxies and the enhancement of star formation in the
universe.  Near-infrared spectroscopy is ideal for studying stellar
populations because it traces not only the presence of hot young stars
though near-infrared hydrogen recombination lines, such as the
Brackett $\gamma$ line, but also the presence of red stars,
particularly red supergiants, and the presence of shocks from
supernovae activity.  Red supergiants and giants can be determined
from the presence of various absorption features in the H- and
K-bands, such as the CO and Si~{\small I} absorption features in the
H-band and the CO bands longward of 2.29~$\mu$m \citep{omo93}.
Supernova activity can be inferred from the presence of Fe~{\small II}
emission lines associated with the supernova activity \citep{omd89,
getal91, c93, fw93, getal97, mds02, aetal03} and possibly even
molecular hydrogen line emission \citep{omd89, getal91, fw93}
(although molecular hydrogen line emission may be produced by other
sources; see \citet{m94} and \citet{vr97}). These diagnostics can be
used together to characterize the different ages of the various
stellar populations within galactic nuclei.

Once the nuclear stellar populations of spiral galaxies are well
defined, they can be used to answer a number of specific scientific
questions.  To begin with, this will illuminate how star formation
proceeds in the nuclei of ``normal'' spiral galaxies, and whether that
star formation is either continuous or a series of short bursts.
Once we understand the functionality of star formation over time, we
can determine how the nuclei of spiral galaxies have evolved.  We can
also determine how morphological features, particularly bars, or how
environmental influences, such as interactions, influence nuclear star
formation activity.  Additionally, we can probe the link between star
formation activity and Seyfert or LINER activity.

Much of the work at near-infrared wavelengths has focues on unusual
objects with either AGN or very strong star formation activity.
Recent surveys include surveys of ultraluminous infrared galaxies
\citep{getal95, metal99, msmetal01, bwd01}, luminous infrared galaxies
\citep{gjdetal97}, starbursts \citep{e97, i00, cdd01}, Seyferts
\citep{i00, stl01, rkp02, betal02}, LINERs \citep{letal98, aetal00,
stl01}, and interacting galaxies \citep{var98}.  However, relatively
little near-infrared spectroscopic survey has been done with
``normal'' nearby spiral galaxies (see \citet{mbpetal01} for an
example).  Such surveys are necessary, though, for understanding
nuclear star formation histories, for identifying the triggers that
enhance nuclear star formation, and for setting a baseline for the H-
and K-band emission from normal spiral galaxies.  Without such a
baseline, the contribution of relatively quiescent stellar populations
to H- and K-band emission in starbursts as well as the older stars'
contribution to the overall mass and luminosities in these systems
will remain unknown, and the enhancement in star formation activity in
exotic systems like ultraluminous infrared galaxies will have no
context.

Therefore, we have undertaken a near-infrared spectroscopic survey of
a subset of the galaxies in the ISO Atlas of Bright Spiral Galaxies
(\cite{betal02a}, henceforth referred to as Paper 1) with the purpose
of understanding the nuclear stellar populations and star formation
histories of these objects.  The data includes K-band spectroscopy for
41 galaxies, with additional H-band spectroscopy for 20 galaxies.  We
first discuss in Section~\ref{s_data} the sample, the observations,
and the data processing.  Next, we divide the galaxies into two
groups: quiescent galaxies and non-quiescent galaxies.  In
Section~\ref{s_quies}, the spectra of the quiescent galaxies are
described in detail, combined together to make a composite quiescent
spectrum, and analyzed using population synthesis models.  The
non-quiescent galaxies are also described in detail in
Section~\ref{s_nonquies}, with particular focus on how their nuclear
star formation activity may be related to morphology, environment, and
AGN activity.  After this, we determine the relative fractions of
stars formed from starbursts in these systems.  Finally, in
Section~\ref{s_specphotcomp}, we compare quiescent and non-quiescent
galaxies in plots of the $\frac{f_{12\mu m}}{f_K}$ ratio to the
$f_{12\mu m}$ or $f_K$ luminosities for the inner $15\arcsec$ as an
effort to justify using mid-infrared fluxes normalized by
K-band fluxes (used in \citet{betal02b}, henceforth referred to as Paper 2) 
to trace star formation activity.

\section{Data \label{s_data}}

\subsection{Sample}

The galaxies in this ISO Bright Galaxies Project sample are a subset
of a complete, magnitude-limited set of galaxies selected from the
Revised Shapley-Ames (RSA) Catalog \citep{st87}. The sample comprised
galaxies with Hubble types between S0 and Sd and with magnitudes
B$_T$~=~12 or brighter; galaxies in the Virgo Cluster were excluded.
A randomly selected, subset of these galaxies was observed by ISO
based on target visibility.  This produced a total of 77 galaxies that
are representative of the range of Hubble types in the RSA Catalog.
Detailed information on the sample is presented in Paper 1.

Of these 77 galaxies, we observed all targets that could be feasably
observed with the the CGS4 infrared spectrometer at the United Kingdom
Infrared Telescope as well as some additional targets that were
observable with the SPEX near-infrared spectrometer at the NASA
Infrared Telescope Facility \citep{retal03}.  This presented two
contraints on the subsample.  The first was on the sensitivity of the
instruments.  We were effectively limited to working with targets
where the continuum could be detected above the sky noise within 2~min
because of the timescale of the variability of atmospheric emission
lines that produced the background emission in our spectra.  The other
contraint was created by the declination range of the telescopes used.
All observed targets had to fall within the declination range of $-51
< \delta < +66$.  In the end, a subset of 41 galaxies in K-band and 20
galaxies in H-band were observed with one of the spectrometers with
sufficient signal-to-noise in their spectra for analysis.  Details on
the morphology, distances, and nuclear activity of this subsample are
given in Tabel~\ref{t_sample}.

\subsection{Observations}

Details on specific information about the observations of individual
galaxies are given in Tables~\ref{t_obsinfok} and \ref{t_obsinfoh}.
General information on the observations with each specific instrument
is given below.

\subsubsection{SPEX}

SPEX observations were made on UT dates 5 - 8~June 2000; 16, 17, and
19~September~2000; and 4 - 5~April~2001.  Each target was observed
with the instrument in the short wavelength cross-dispersed mode,
which covered a wavelength range from 0.8 to 2.5~$\mu$m (although we
only use the 1.2 - 2.5~$\mu$m range).  The slit used was
0.5~$\times$~15~$\arcsec$, giving a spectral resolution of 1200.  The
galaxies' nuclei (in the J- or K-band) were centered in the slit and
used for guiding in all cases.  The observations consist of a series
of alternating 2 min observations of the galaxies and 2 min
observations of an empty region of sky $\sim$2$\arcmin$ off the
source, with total on-source integration times of 20 - 30~min.  Argon
lamp spectra were taken for wavelength calibration, and quartz
tungsten halogen (QTH) lamp spectra were taken for flatfielding the
spectra.  An A0V star was observed as a spectroscopic standard for all
targets, and, for targets observed in September 2000 and April 2001, a
G type star was also observed as a spectroscopic standard.  The
spectroscopic standards were selected to be as close to the target
galaxies as possible, preferably within 10$^\circ$.  These stars were
observed using the same configuration and methods that the galaxies
were observed with except that the integration times were shorter and
that the stars observed in September 2000 and April 2001 were observed
by nodding along the slit rather than completely off-source.

\subsubsection{CGS4}

CGS H- and K-band spectroscopic observations were made on UT dates 7 -
10 May 2001.  The observations were taken with the 40 lines mm$^{-1}$
grating, and the slit used was 0.5~$\times$~15~$\arcsec$, giving a
spectral resolution of $\sim$800.  The galaxies' nuclei (in H- and
K-band) were centered in the slit in all cases, and off-source point
sources were used for guiding.  The observations consist of a series
of alternating 2 min observations of the galaxies and 2 min
observations of an empty region of sky $\sim$2$\arcmin$ off the
source, with total on-source integration times of 20~min.  Bias frames
were taken for removing systematic readout effects, Argon lamp spectra
were taken for wavelength calibration, and incandescent lamp spectra
were taken for flatfielding the spectra.  For each galaxy, a G type
star was observed as a spectroscopic standard.  The spectroscopic
standards were selected to be as close to the target galaxies as
possible, preferably within 10$^\circ$.  These stars were observed
using the same configuration and methods that the galaxies were
observed with except that the integration times were shorted and that
the stars were observed by nodding along the slit as opposed to
completely off-source.

\subsection{Data Reduction}

\subsubsection{SPEX}

First, the background frames were subtracted from the target frames.
Next, the frames were divided by the QTH lamp frames.  The individual
orders were extracted from each frame, and spectra of the central
sources were extracted from the H and K orders.  The spectra were
wavelength calibrated using Argon arc lamp spectra.  The slopes of the
spectra were corrected to account for small changes in the waveband
covered per pixel.  Next, the spectra from each frame were normalized
relative to each other by performing a least-squares fit to solve for
the scalar multiplier that minimized the differences between the first
spectrum and subsequent specta.  For each wavelength value, the data
were then combined in an iterative process where data points that
deviated by more than 3 standard deviations from the median were
removed until all data points were within three standard deviations of
the median; the median and standard deviation of the mean were
then taken as the flux density measurement and the error at that
wavelength.  Finally, the spectra were smoothed using a Gaussian
function.

The spectroscopic standard was processed the same way, except that the
blackbody emission was divided out and the hydrogen recombination
lines were removed using one of two possible methods.  In some cases,
the lines were simply removed by fitting the sums of a Voigt function
and a continuum to the region with the line, then adding the Voigt
function to the spectra.  This was most effective for removing the
Brackett-$\gamma$ lines from A0V and G spectroscopic standards.  Other
lines were removed by dividing the spectra by the spectra of
previously observed stars of the same type.  This was most effective
for removing higher order Brackett lines from the H-band spectra of
A0V standards (using data from \citet{metal98}).  If both A0V and G
type stars had been observed, then a spectroscopic standard spectrum
was made by using the A0V spectrum but by replacing the region around
the Brackett-$\gamma$ line with the spectrum from the G type star.
Otherwise, the full spectrum of the A0V star was used as the
spectroscopic standard spectrum.  Once all the spectra were ready, the
galaxy spectra were then divided by their corresponding spectroscopic
standard spectra.

\subsubsection{CGS4}

Data reduction for CGS4 data followed the data processing for SPEX
data with some exceptions.  First, only one linear order was present
in the spectra, so no work needed to be done extracting the orders.
Second, since only G type spectroscopic standards were observed, only
the Brackett-$\gamma$ line needed to be corrected in the K-band; the
hydrogen recombination lines were so weak in the H-band spectra that
no corrections needed to be made.  Third, the sampling of the spectra
performed during the observations produced a step pattern between odd
and even pixels that needed to be removed from the individual
spectroscopic standard spectra before they were median combined.

\section{Quiescent Spectra \label{s_quies}}

33 of the 41 nuclear K-band spectra we examined had nearly identical spectra 
with identical qualitative characteristics:

- The spectra all have similar slopes, at least from 2.10~$\mu$m
to the edge of the fist CO band at 2.29~$\mu$m.

- The spectra all contain strong CO absorption lines that are 
characteristically found in K and M stars (see examples in \citet{kh86}, 
\citet{lr92}, \citet{omo93}, \citet{wh97}, and \citet{f00}).

- The spectra all contain various metal absorption features, most notably
the Na {\small I} lines at 2.206 and 2.209~$\mu$m and the Ca {\small I} 
lines at 2.261, 2.263, and 2.266~$\mu$m, that are most prominent
in K and M stars (see examples in \citet{kh86}, \citet{wh97}, 
and \citet{f00}).

- No Brackett-$\gamma$ emission lines are evident in 
the spectra.  Although ionized gas may be present, the negligible 
recombination line flux in the K-band implies relatively few young stars 
and therefore relatively little recent star formation activity.

- No H$_2$ lines are evident in the spectra.  If H$_2$ line emission is
interpreted as originating from shocks connected to supernova activity, 
then the lack of these lines in the spectra of these galaxies
implies that nuclear supernova activity is relatively weak or nonexistent.

The H-band spectra for 15 of these 33 ``quiescent'' spectra share similar
characteristics to the K-band spectra:

- The spectra al have similar slopes within the 1.55 - 1.75 $\mu$m wavelength
range.

-  The spectra all contain strongly defined CO absorption bands that are
characteristically found in K and M stars (see examples in \citet{lr92},
\citet{omo93}, \citet{dbj96}, and \citet{metal98}).

- The spectra contain metal absorption lines, such as the 1.589~$\mu$m 
Si~{\small I} line and the 
1.711~$\mu$m Mg {\small I} line, that are characteristically found in cooler 
stars (see \citet{omo93} and \citet{metal98} for examples).

- The spectra contain no Fe~{\small II} 1.644~$\mu$m line emission, 
which is associated with supernova activity.

\begin{figure}
\plotone{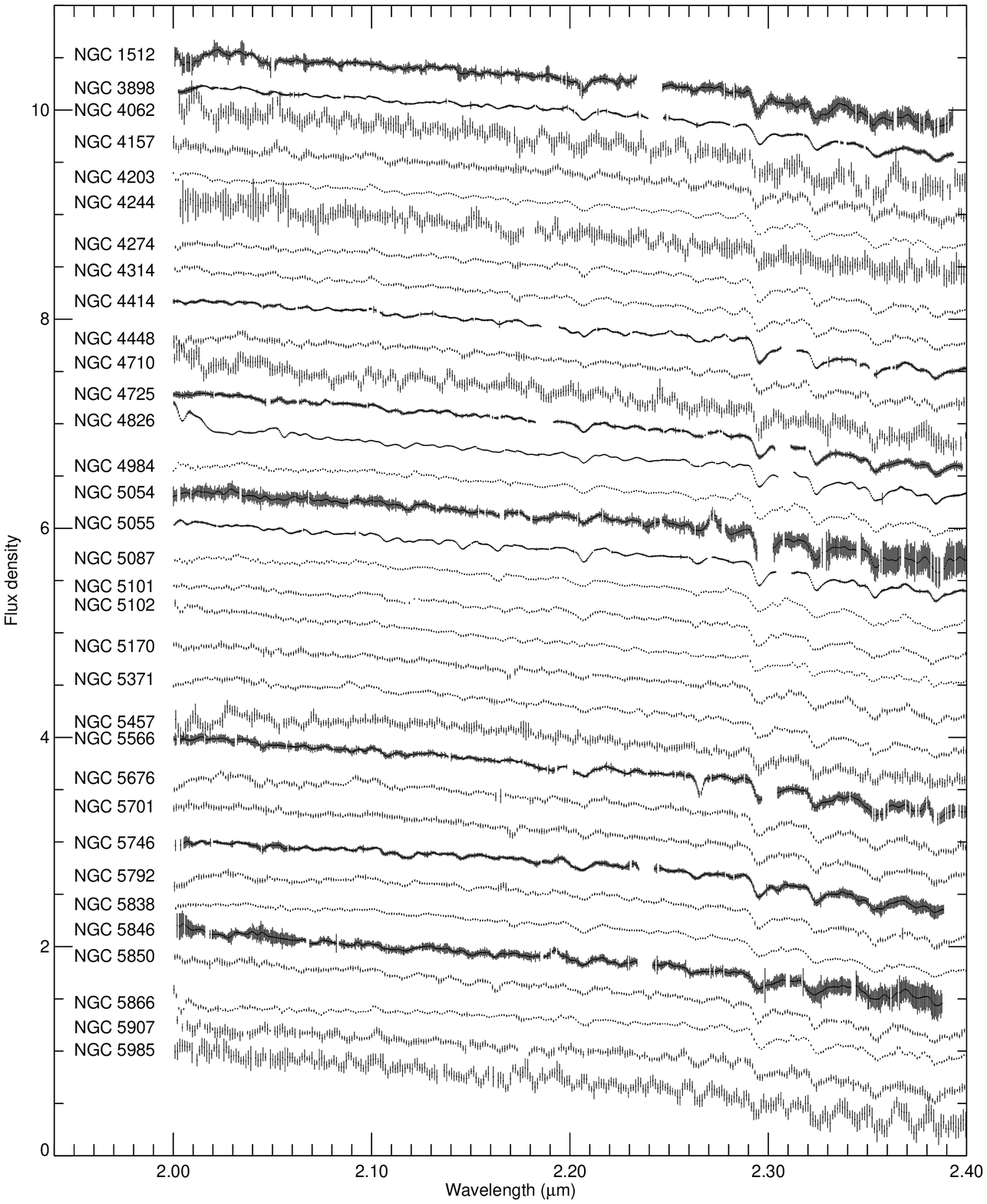}
\caption{The K-band spectra of the 33 quiescent galaxies.  The flux density
units are arbitrary but are in terms of flux per unit wavelength.  An
artifical offset has been applied to display all the spectra in this plot.}
\label{f_allquiesspeck}
\end{figure}

\begin{figure}
\plotone{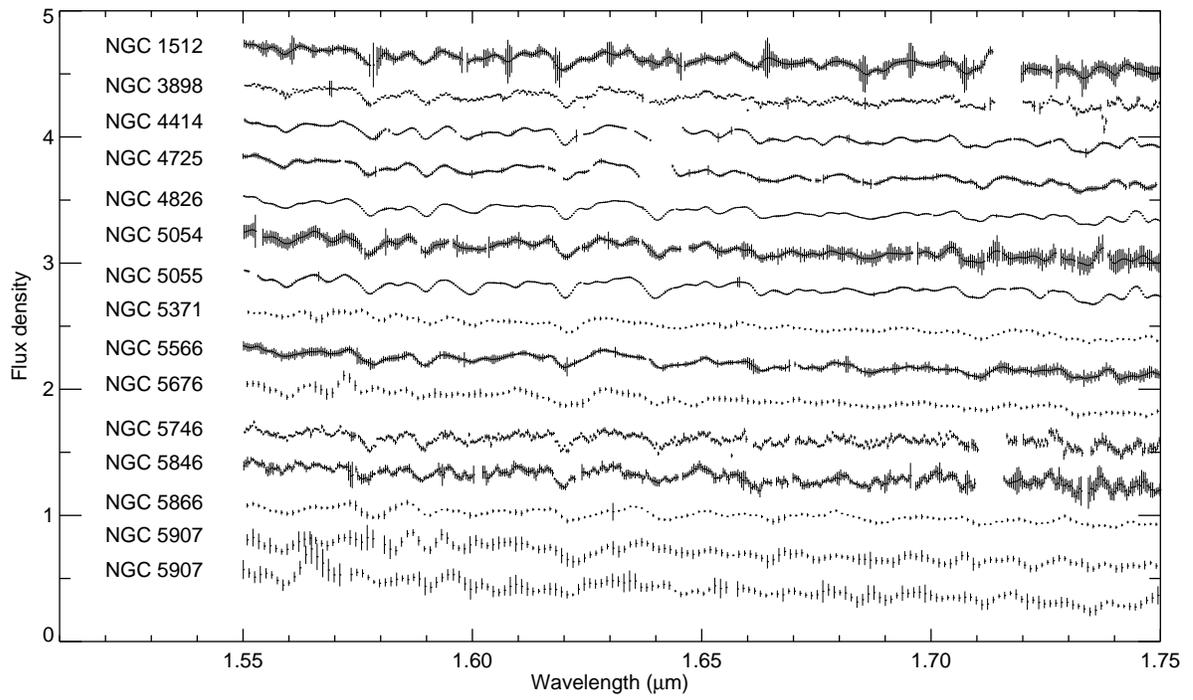}
\caption{The 15 available H-band spectra of quiescent galaxies.  
The flux density units are arbitrary but are in terms of flux per unit 
wavelength.  An artifical offset has been applied to display all the 
spectra in this plot.}
\label{f_allquiesspech}
\end{figure}

We display these 33 ``quiescent'' K-band spectra in 
Figure~\ref{f_allquiesspeck} and the 
corresponding ``quiescent'' H-band spectra in Figure~\ref{f_allquiesspech}.  
The uniformity of
the spectra is quite striking, especially when they are compared to
samples of other classes of objects.  The LIRGs studied in \citet{gjdetal97}
or the interacting galaxies studied in \citet{var98}, 
for example, have variable spectral slopes, and the strengths of
Brackett-$\gamma$ and H$_2$ lines are variable.  In contrast, the slopes of
the spectra of these quiescent galaxies are uniform, and the relative strengths
of absorption features is consistent throughout the group.

\subsection{A Composite Quiescent Spectrum}

Based on the spectra of these objects, it is very clear that the
K-band emission from these objects is dominated entirely by old red
stars that dominate the central bulges of these galaxies.  We see such
a large collection of spectra as an excellent opportunity.  By
combining these spectra together, we can create a composite
``quiescent'' spectrum.  This quiescent spectrum could then be treated
as representing an old red stellar component in the spectra of
galaxies where either continuum emission from an AGN source or hot
dust or emission lines from young stars or supernovae or are present.

To build a quiescent K-band spectrum, we began by selecting the 32 of
the 33 galaxies above that were S0 or spiral galaxies.  (NGC~5846 is
classified by the Third Reference Catalogue of Bright Galaxies
\citep{detal91} (RC3) as an elliptical galaxy and was therefore left
out of this analysis.  Composite spectra made from only S0s or only
spiral galaxies differed a statistically insignificant amount from
each other, so we used the S0 and spiral galaxies together.)  We first
redshift-corrected all of the spectra to the rest frame.  Next, we
finely resampled the spectra from 2.0~$\mu$m to 2.4~$\mu$m at
intervals of $10^{-4}$~$\mu$m using a spline procedure.  We then
normalized the spectra and median combined them using the same method
we used to normalize and combine the spectra for individual galaxies.
A similar procedure was followed to produce quiescent H-band spectra
except that the 14 spectra that were used were resampled from
1.55~$\mu$m to 1.75~$\mu$m at intervals of $10^{-4}$~$\mu$m.

\begin{figure}
\plotone{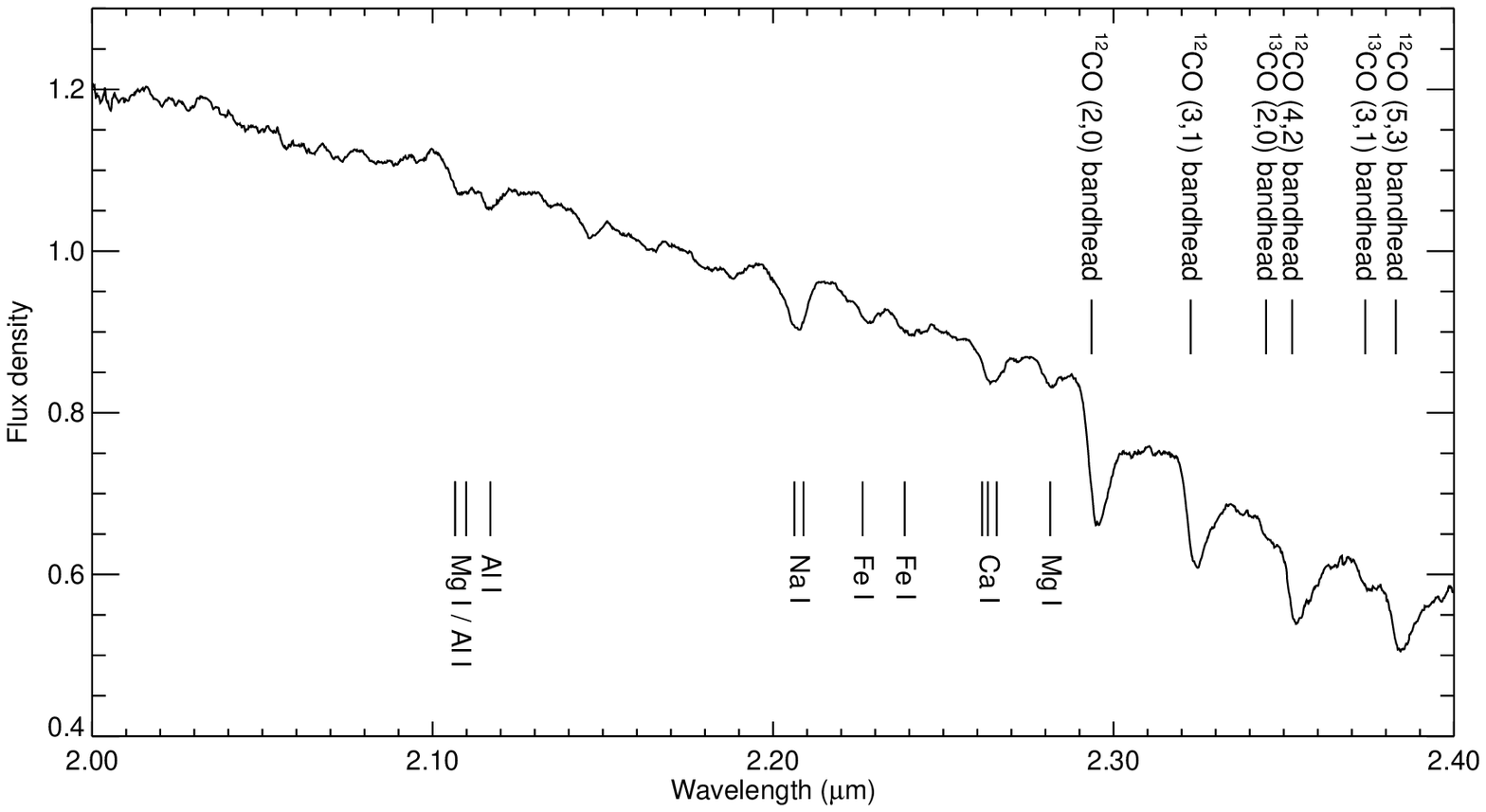}
\caption{The composite quiescent spectrum in the K-band.  The flux density
units are arbitrary but are in terms of flux per unit wavelength.  The median 
value of the flux density in the region between 2.10 and 2.28~$\mu$m has been 
normalized to 1.  Strong absorption features have been marked.}
\label{f_quiesspectrumk}
\end{figure}

\begin{figure}
\plotone{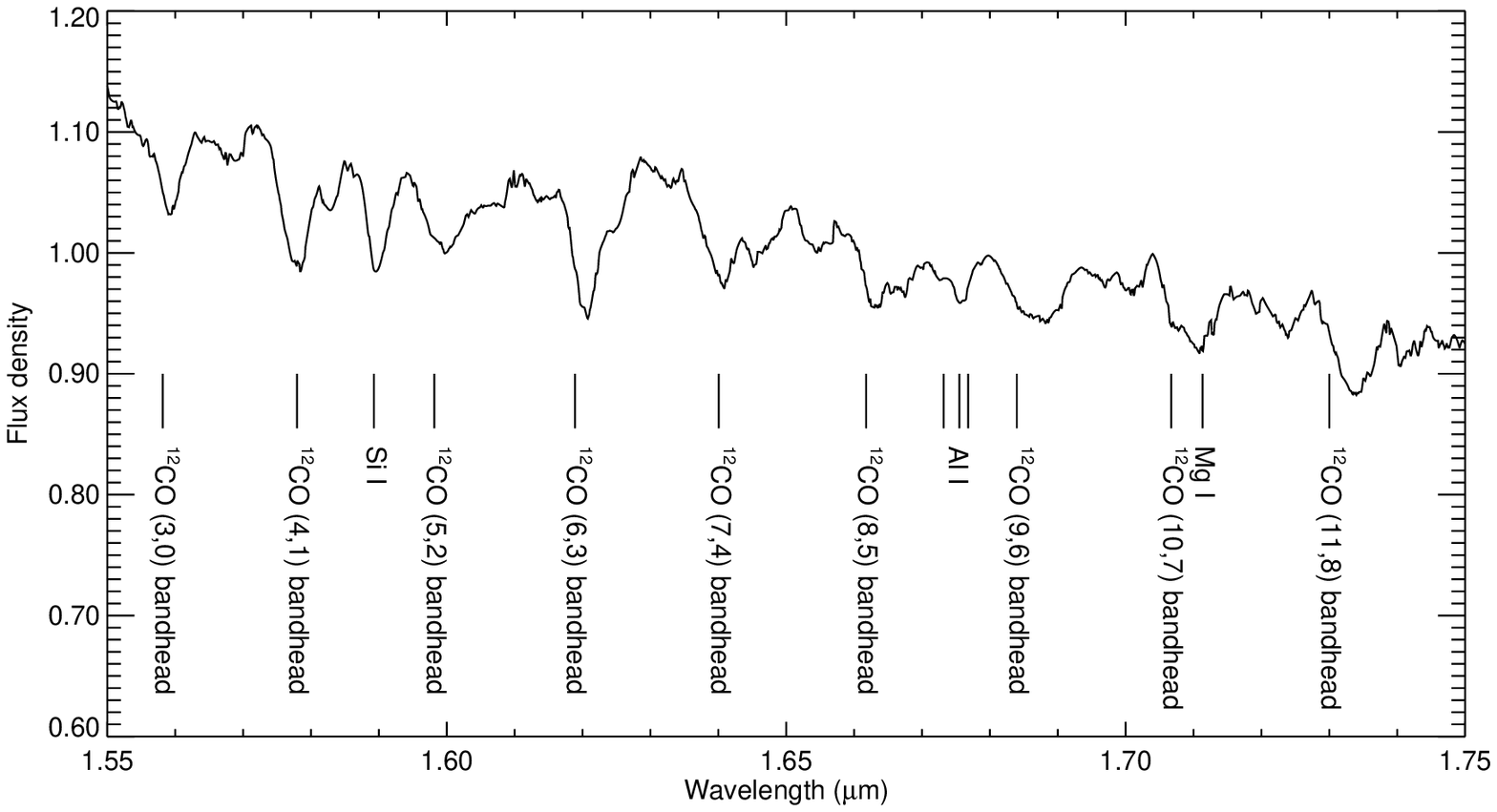}
\caption{The composite quiescent spectrum in the H-band.  The flux density
units are arbitrary but are in terms of flux per unit wavelength.  The median 
value of the flux density in the region between 1.55 and 1.75~$\mu$m has been 
normalized to 1.  Strong absorption features have been marked.}
\label{f_quiesspectrumh}
\end{figure}

We show the resulting quiescent spectra in Figure~\ref{f_quiesspectrumk}
and Figure~\ref{f_quiesspectrumh}.
Spectral features, namely the metal absorption lines and the CO bandheads,
were identified using spectral line data from \citet{kh86}, \citet{wl92},
and \citet{g94}.  Tables~\ref{t_quieskline} and \ref{t_quieshline} 
list the identified features.

\subsection{Analysis with Population Synthesis \label{s_quies_synth}}

To better understand the composite spectrum, we will analyze
measurements of features in the spectrum using the Starburst99
\citep{letal99} population synthesis models.  Starburst99 only
considers two star formation scenarios: an instantaneous burst of star
formation, and continuous star formation.  We will consider
both scenarios in our analysis.  We used most of the default input
choices for Starburst99 with the exception of the initial mass
function (IMF).  We used solar metallicities because we assume that
the galaxies used to build the composite quiescent spectrum should
have similar metallicities to the Milky Way.  Although we will
primarily focus on model results that depend on using a Salpeter IMF
with a slope of 2.35 and an upper mass cutoff of 100~M$_\odot$, we
will also briefly mention the implications of using either an IMF with
a slope of 2.35 and an upper mass cutoff of 30 M$_\odot$ or an IMF
with a slope of 3.30 and an upper mass cutoff of 100 M$_\odot$.

Several different near-infrared spectroscopic indicators are modeled in
Starburst99, but only a few of these indicators trace older stellar
populations.  We will use Starburst99's predictions of the equivalent widths
of the Si 1.59~$\mu$m line, the CO~6-3 feature at 
1.62~$\mu$m, and the CO~2-0 feature at 2.29~$\mu$m (\citet{omo93} describe
the motivations for using these lines in near-infrared spectral analysis).
Table~\ref{t_quiesmeas} list measurements of these 
features using the ``line'' and continuum locations described in 
\citet{omo93}.  Errors in the line
measurements are dominated not by random errors in the measurements of the
spectral lines but in systematic errors related to the wavelength calibration
of the spectrum.

\subsubsection{The Instantaneous Burst Scenario}

\begin{figure}
\plottwo{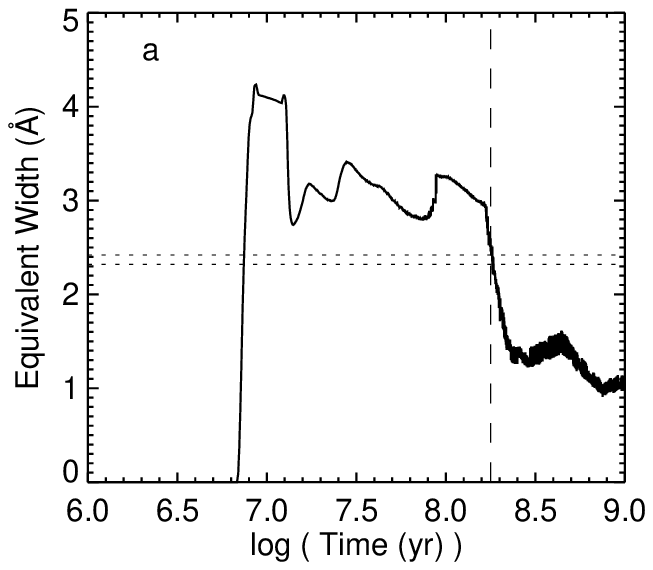}{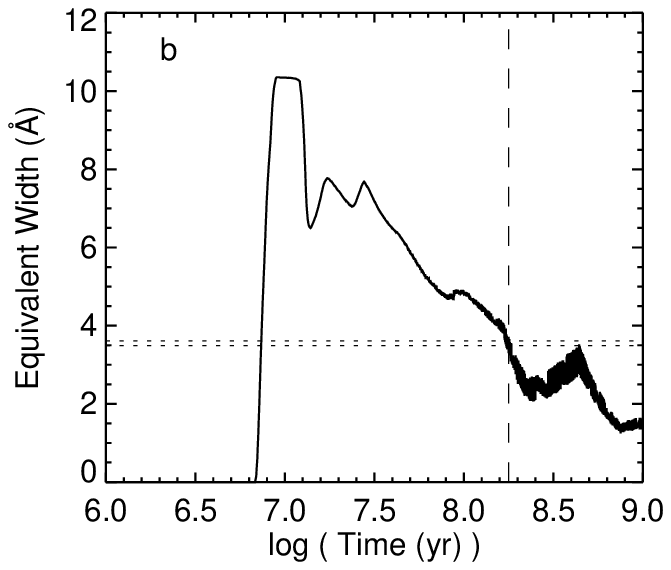}
\plotone{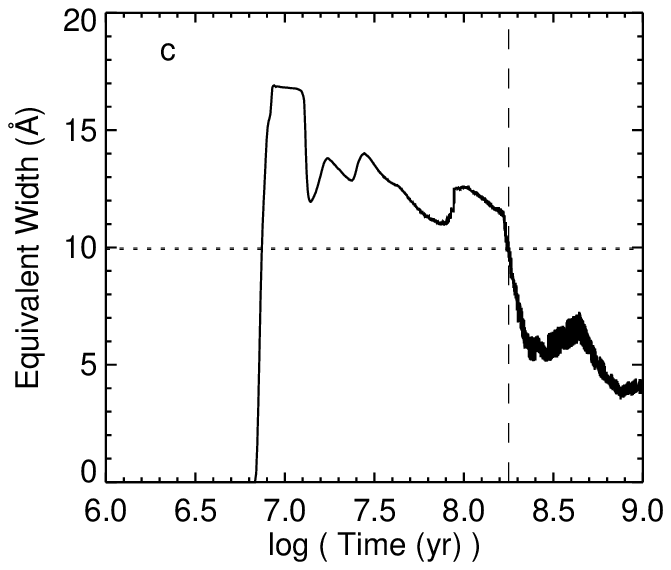}
\caption{Plots of (a) the Si~{\small I} 1.59~$\mu$m equivalent width,
(b) the CO 6-3 1.62~$\mu$m equivalent width, and (c) the CO 2-0
2.29~$\mu$m equivalent width after an instantaneous burst of star
formation that produces an IMF with a slope of 2.35 and an upper mass
cutoff of 100~M$_\odot$, as determined using Starburst99.  The
$1\sigma$ errors in the measured equivalent widths for the composite
quiescent spectrum are superimposed as dotted horizontal lines.  The
dashed vertical lines shows where the $\chi^2$ value is at a minimum.}
\label{f_s99inst}
\end{figure}

Figure~\ref{f_s99inst} shows plots of Starburst99's predictions of the 
equivalent widths over time after an instantaneous burst of star formation 
that produces a Salpeter IMF.  The spectral line measurements and errors, 
overlaid as horizontal lines, all cross the model at two locations in these 
plots.  The most representative times have been determined quantitatively
by searching for the location where the $\chi^2$ value is minimized. 
The $\chi^2$ value for time t is given by 
\begin{equation}
\chi(t)^2 = \sum_{i} \frac{(w_{meas,i} - w(t)_{model,i})^2}{\sigma_{meas,i}^2}
\end{equation}
where w$_{meas,i}$ is the measured equivalent width for line i, 
$\sigma_{meas,i}$ is the error in the measurement, and w(t)$_{model,i}$ is
the modeled equivalent width at time t.

The first 
location at $\sim$7~Myr seems less likely to be representative of the
stellar populations.  At this time after the burst, the bluest stars on the 
main sequence are beginning to evolve into red supergiants, which have 
atmospheres dominated by molecular and metal absorption lines.  However, 
as massive stars are evolving past the red supergiant phase, supernova 
activity is quite strong at this point in time, which implies
that shock excitation lines such as the Fe~{\small II} and H$_2$ emission
lines should be quite strong in the composite quiescent spectrum, which is not
the case.  Furthermore, the stellar populations are evolving rapidly, which
would suggest much more variation in the spectral features than what is found
among the quiescent galaxies.

The other location at $\sim$180~Myr seems to be more representative of the
nuclear stellar populations.  At this stage in stellar evolution, most
massive stars have completely evolved and no longer significantly 
contribute to the H- or K-band emission in these objects.  Neither supernovae
nor ionizing stars are present, so no strong emission lines are expected.
What is left is an old, red stellar population that will produce continuum
emission that includes metal and molecular absorption features.  \citet{oo00}
do caution that interpretation of stellar populations older than 100~Myr,
particularly using near-infrared absorption features, is problematic because 
of difficulty in modeling asymtotic giant branch stars.
We will therefore treat the Starburst99 output as a rough characterization of
the actual stellar populations.

Given the approximate age of the system, it is possible to estimate the
ratio of K-band luminosity to the mass of the system for this particular case.
From Starburst99, a $10^6$~M$_\odot$ system will lose
$2.3 \times 10^5$~M$_\odot$ of its mass through stellar winds and 
supernovae, thus leaving $7.7 \times 10^5$~M$_\odot$ in stars.
The K-band absolute magnitude at this time is $-13.7$, which corresponds to
a power of $6.1 \times 10^{31}$~W.
This gives a K-band luminosity-to-mass conversion factor of 
$1.3 \times 10^{-26}$~M$_\odot$~W$^{-1}$.

The other IMFs considered here, which are weighted more strongly
towards low mass stars, do produce results that differ from the
Salpeter IMF.  The age of the stellar population does not change when
a smaller upper mass cutoff is used and changes an insignificant
amount if an IMF with a steeper slope is used.  In these scenarios,
however, less of the total initial mass is lost through supernovae or
stellar winds because less mass is in stars that would have high mass
loss through stellar winds or that would undergo supernovae
explosions.  This results in the luminosity-to-mass conversion factor
varying somewhat among the three IMF scenarios, but it does not vary
by orders of magnitude.

In summary, the results from the Starburst99 simulations demonstrate that
an old stellar population formed in an instantaneous burst can reproduce
the observed composite quiescent spectrum as well as the spectra of the
galaxies used to build the quiescent spectrum.  Changing the output IMF in 
Starburst99 will still lead to a stellar population that can reproduce the 
measured absorption line features. 

\subsubsection{The Continuous Star Formation Scenario}

\begin{figure}
\plottwo{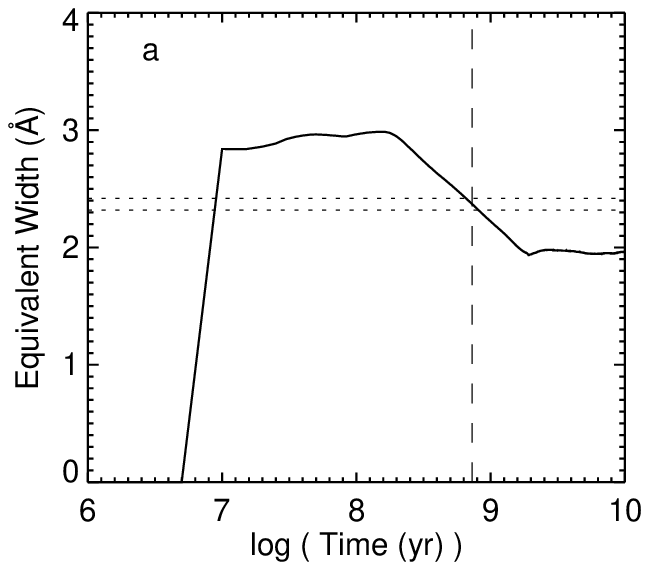}{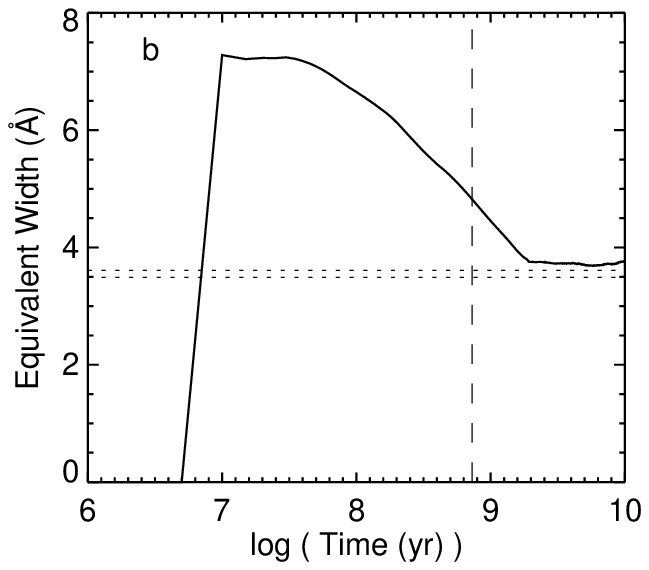}
\plotone{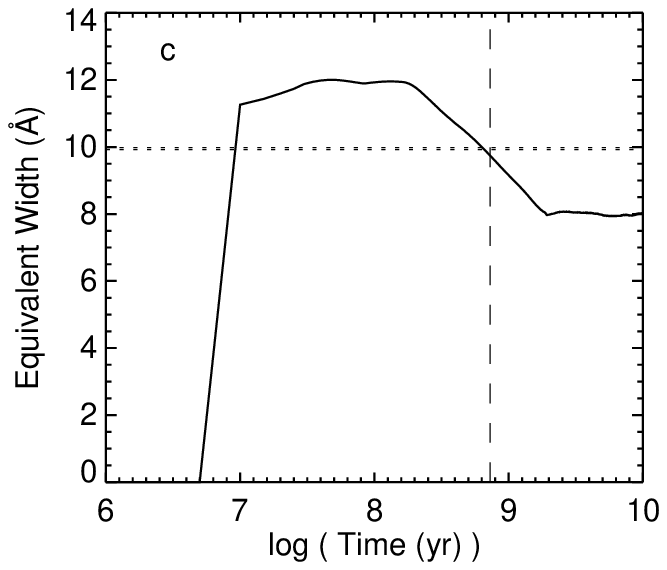}
\caption{Plots of (a) the Si~{\small I} 1.59~$\mu$m equivalent width,
(b) the CO 6-3 1.62~$\mu$m equivalent width, and (c) the CO 2-0
2.29~$\mu$m equivalent width after the onset of continuous star
formation that produces an IMF with a slope of 2.35 and an upper mass
cutoff of 100~M$_\odot$, as determined using Starburst99.  The
$1\sigma$ errors in the measured equivalent widths for the composite
quiescent spectrum are superimposed as dotted horizontal lines.  The
dashed vertical lines shows where the $\chi^2$ value is at a minimum.}
\label{f_s99cont}
\end{figure}

Figure~\ref{f_s99cont} shows plots of Starburst99's predictions of the 
equivalent widths over time after the onset of continuous star formation 
that produces a Salpeter IMF.  The spectral line measurements and errors, 
overlaid as horizontal lines, all cross the model at two locations in these 
plots.  The earlier crossing, at a time of $\sim$10~Myr, seems an 
unreasonable alternative for reasons similar for those given above for the
earlier time in the instantaneous burst models.  This leaves the later time
as a given alternative.  In this case, the age of the stellar population is
$\sim$730~Myr.  The reason for this greater age than in the instantaneous
burst scenario is that the continuous star formation continuously adds red 
supergiants with absorption features.  Therefore, more time is needed for
stars with shallower absorption features to dominate the spectrum and match
these measurements.

For a few reasons, the continuous star formation scenario is more
problematic than the instantaneous burst scenario.  First of all, the
Si, CO 2-0, and CO 6-3 equivalent widths do not all correspond to
similar times after the onset of star formation.  The Si and CO 2-0
equivalent widths correspond to an age of 600 - 800~Myr, whereas the
CO 6-3 equivalent width corresponds to an age of 2~Gyr or later.  The
discrepancy between the two measurements suggests that the continuous
star formation scenario provides the wrong mix of K and M stars to
reproduce the stellar population.  More importantly, however, the
continuous star formation scenario predicts a Brackett-$\gamma$
equivalent width of $\sim$30~\AA \ at $\sim$730~Myr.  No
Brackett-$\gamma$ feature within a magniture of being that strong is
evident in the composite quiescent spectrum.  Furthermore, no shock
excitation lines are observed, even though supernova activity is
continuous in this scenario.

Using other IMFs does not solve the problem.  Using a steeper IMF or
decreasing the upper mass cutoff does improve the match between the
measured and modeled absorption features, but the discrepancy among
the best corresponding times for each equivalent width measurement
persists.  Moreover, the other IMF model results still suggest
detectable Brackett-$\gamma$ equivalent widths.  We conclude that
changing the IMF slightly improves the match between the Starburst99
output and the measurements, but the resulting simulated stellar
population still cannot plausibly represent the observed spectra.

\subsubsection{Conclusions from the Analysis with Population Synthesis} 

Ultimately, the data suggests that the instantaneous burst scenario
can reproduce the spectra we have observed but the continuous star
formation scenario cannot.  This seems to be consistant with previous
studies of nuclear star formation as well \citep{k_rc98}.  We will
therefore treat the stellar populations of the nuclei of these
quiescent galaxies as though they formed in an instantaneous burst
180~Myr ago.  

However, these results should be used cautiously.  The results should
not be interpreted as indicating that all of the stars in the
quiescent galaxies' nuclei formed in one burst of star formation
180~Myr ago.  Galaxies are known to be older than 180~Myr, and the
star formation histories are expected to be more complex.  The age
should also not be interpreted as an average age, since the line
widths do not vary linearly with time.  Instead, the age should be
treated as a representative value that can be used with the
Starburst99 models to approximate the stellar populations that
predominate the near-infrared emission.  Future refinements in
modeling near-infrared absorption features with population synthesis
models, the inclusion of more absorption features in the models, and
advances in modeling stellar populations beyond 100~Myr will improve
the characterization of the ages of the stellar populations in these
quiescent systems.  For now, however, we will work with these results
from Starburst99.

The viability of the instantaneous burst scenario for the quiescent galaxies
suggests that sources with H- and K-band emission lines are comprised of 
a younger starburst embedded within an older stellar population 
that is similar to the quiescent stellar population.  We apply this 
interpretation throughout the following secion.

\section{Non-Quiescent Nuclear Spectra \label{s_nonquies}}

Only the nuclear H- and K-band spectra of eight galaxies exhibited any
emission lines.  Some of these galaxies are systems with complex
stellar structures in their nuclei that we will discuss in detail in
the Section~\ref{s_nonquies_complex}.  The other systems have strong
line emission from pointlike nuclei; their spectra are presented as
they are discussed.
  
Interestingly, not all nuclei and other structures that produce
emission lines have characteristically similar spectra.  We discuss a
few of the qualitative variations below, and we provide
Table~\ref{t_nonquiesmeas} to show how the spectra differ
quantitatively.  In most cases, the underlying continua and absorption
lines for these systems closely matches the quiescent composite
spectra derived above with some minor exceptions in the slope of the
spectra.  This implies that any recent starburst activity lies deeply
buried within an older population of evolved stars.

\subsection{Types of Non-Quiescent Spectra}

\begin{figure}
\plotone{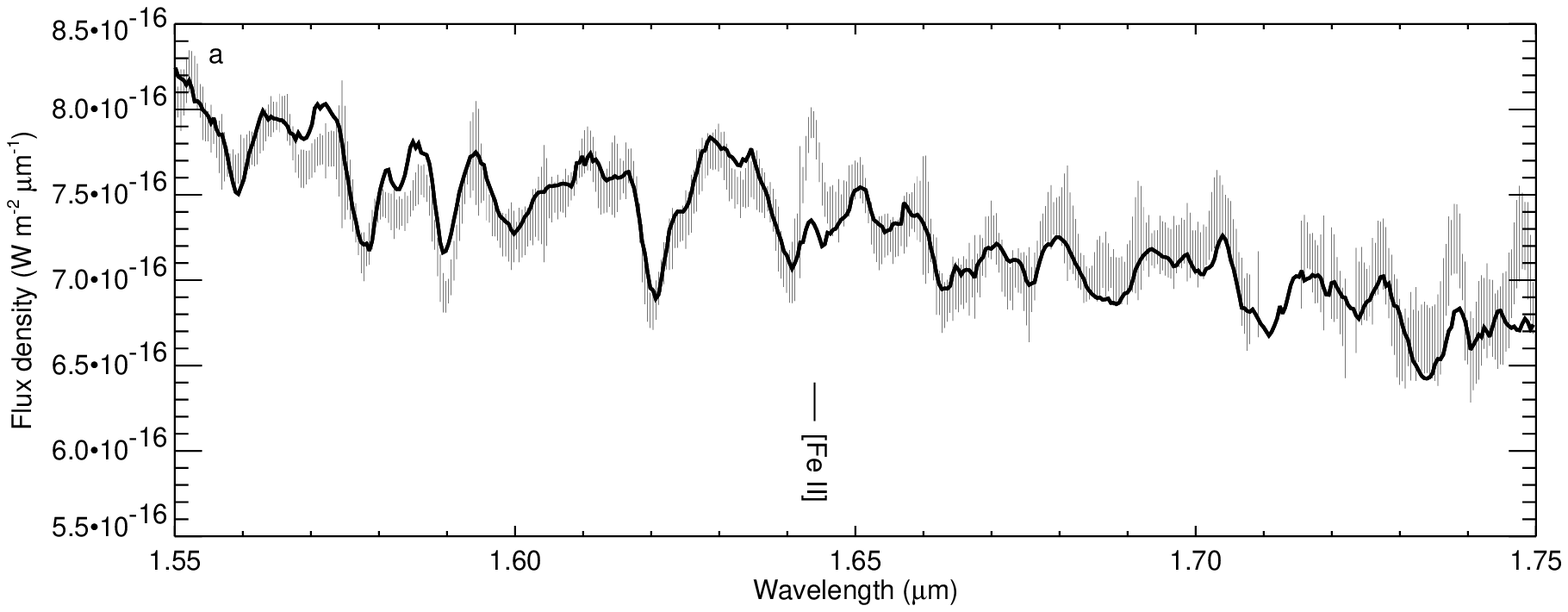}
\plotone{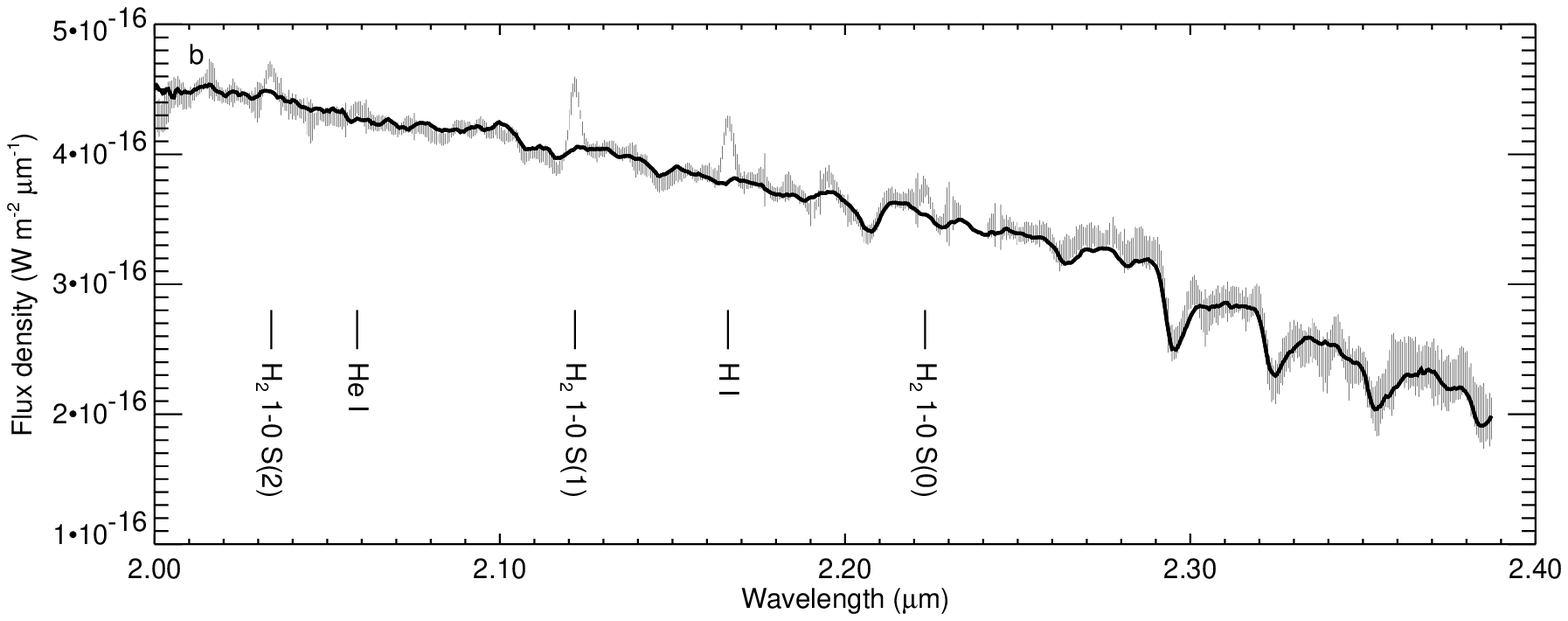}
\caption{The (a) H-band and (b) K-band spectra of NGC~289 (plotted as 
vertical error bars) in the rest frame with the composite quiescent spectrum 
overlaid as a thick black line.  Emission lines of interest are indicated in 
the plot.  All following non-quiescent spectra will be plotted in the same 
format.  Note the strong Fe~{\small II} emission at 1.64~$\mu$m, the strong 
H$_2$ emission at 2.12~$\mu$m and the strong Brackett-$\gamma$
emission at 2.17~$\mu$m.}
\label{f_specn0289}
\end{figure}

\begin{figure}
\plotone{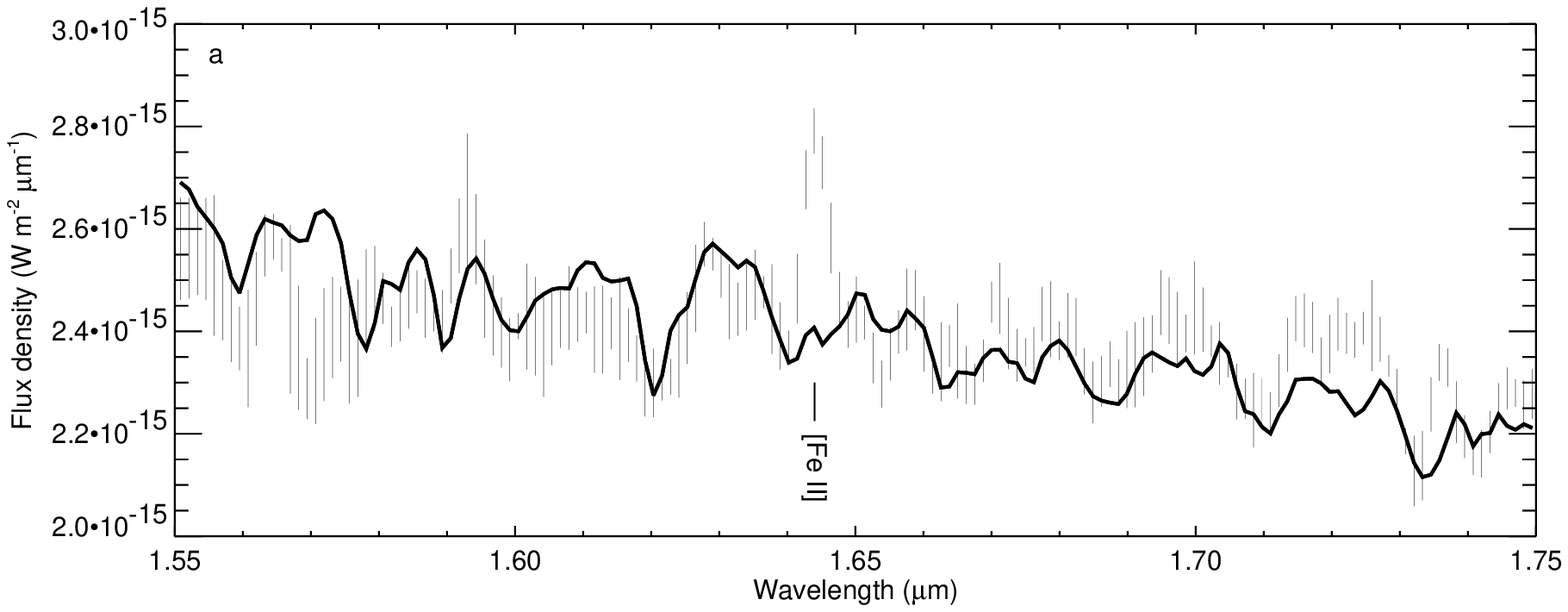}
\plotone{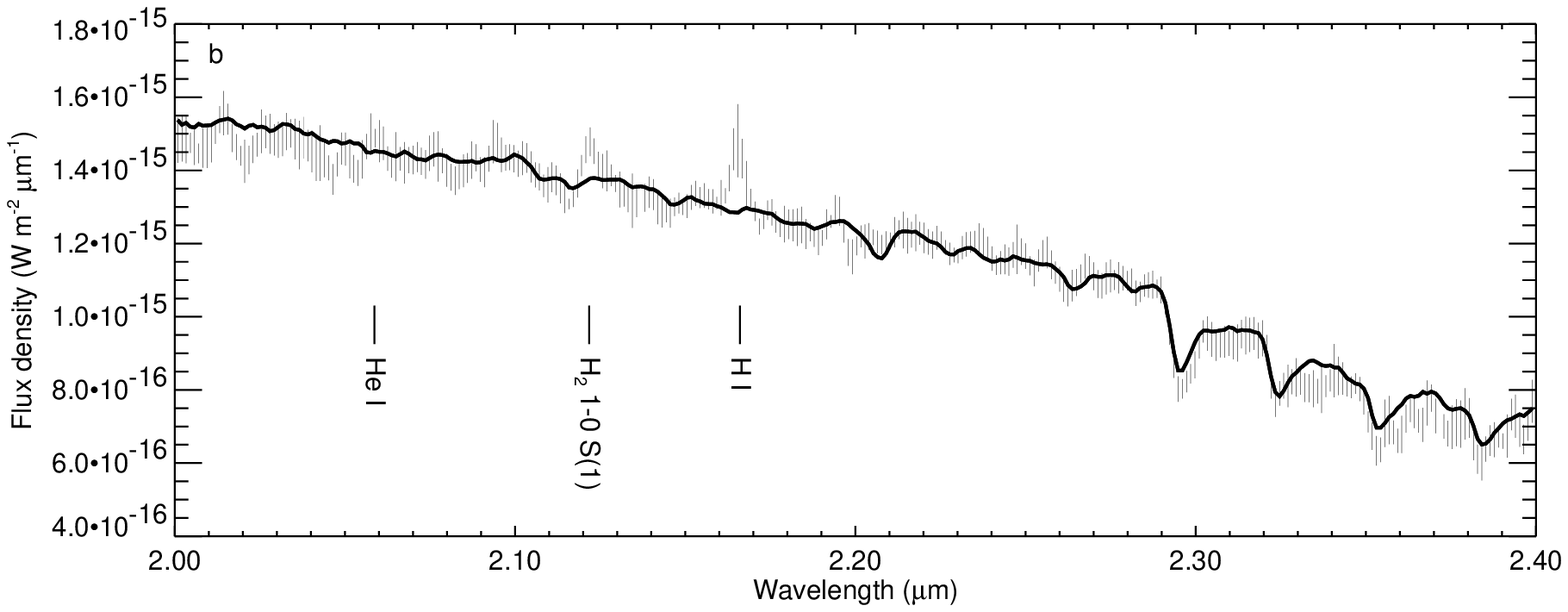}
\caption{The (a) H-band and (b) K-band spectra of NGC~5713.  Again, note
the strong Fe~{\small II}, H$_2$, and Brackett-$\gamma$ emission.}
\label{f_specn5713}
\end{figure}

In NGC~289, NGC~5713, and part of the nuclear regions of NGC~4100,
strong H$_2$ lines, Brackett-$\gamma$ lines, and Fe~{\small II} lines
(when H-band data are available) were detected.  Otherwise, the
galaxies' H- and K-band spectra almost match the quiescent composite
spectra.  Figures~\ref{f_specn0289} and \ref{f_specn5713} show the
spectra for NGC~289 and NGC~5713 with the quiescent H- and K-band
spectra overlaid.  (Spectra and data for NGC~4100 are presented in
Section~\ref{s_nonquies_complex}.)  At this stage in galaxy evolution,
it appears that these systems are seen at a time during which ionizing
stars have not completely evolved off the main sequence but supernovae
are occuring.  According to simulation results from Starburst99, these
conditions are met for an instantaneous burst in the 3.5 - 8~Myr age
range for a Salpeter IMF.

\begin{figure}
\plotone{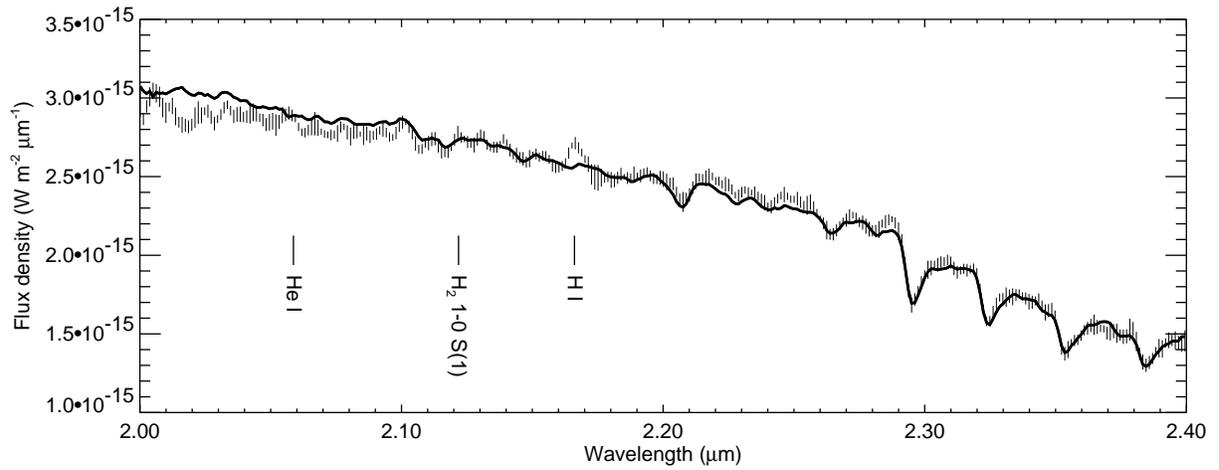}
\caption{The K-band spectrum of NGC~4088. Only Brackett-$\gamma$ is strongly
detected.  (Deviation of the spectrum from the composite quiescent spectrum 
at wavelengths shorter than 2.10~$\mu$m may be a result of the relatively 
high atmospheric variability in that part of the waveband.)}
\label{f_specn4088}
\end{figure}

The A cluster in NGC~1569, part of the nuclear region of NGC~3556, and
the nucleus of NGC~4088 exhibit strong Brackett-$\gamma$ lines but
nondetectable H$_2$ and Fe~{\small II} lines (when H-band data are
available).  He~{\small I} lines may also be present, although it may
be relatively weak compared to the hydrogen emission.  This helium
emission may be a qualitative tracer of O type stars with very strong
winds \citep{k_rp98}.  The spectrum of NGC~4088 is plotted in
Figure~\ref{f_specn4088}.  (NGC~1569 and NCG~3556 are discussed in
more detail in Section~\ref{s_nonquies_complex}.)  The recombination
lines indicate that young ionizing stars are present, but the lack of
shock excitation lines suggests that the stellar population has not
evolved to a point where supernovae are generating shocks in the
interstellar medium.  Therefore, using Starburst99 data, the stellar
population must be less than 3.5~Myr old.

\begin{figure}
\plotone{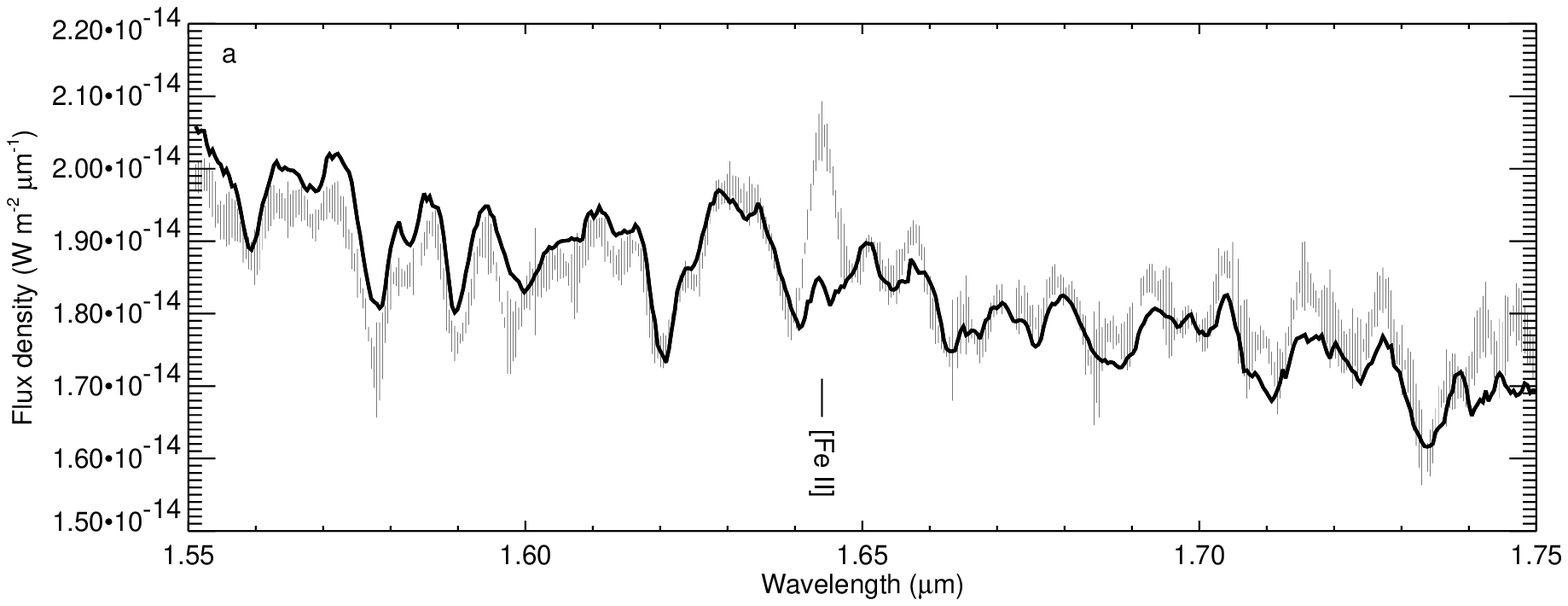}
\plotone{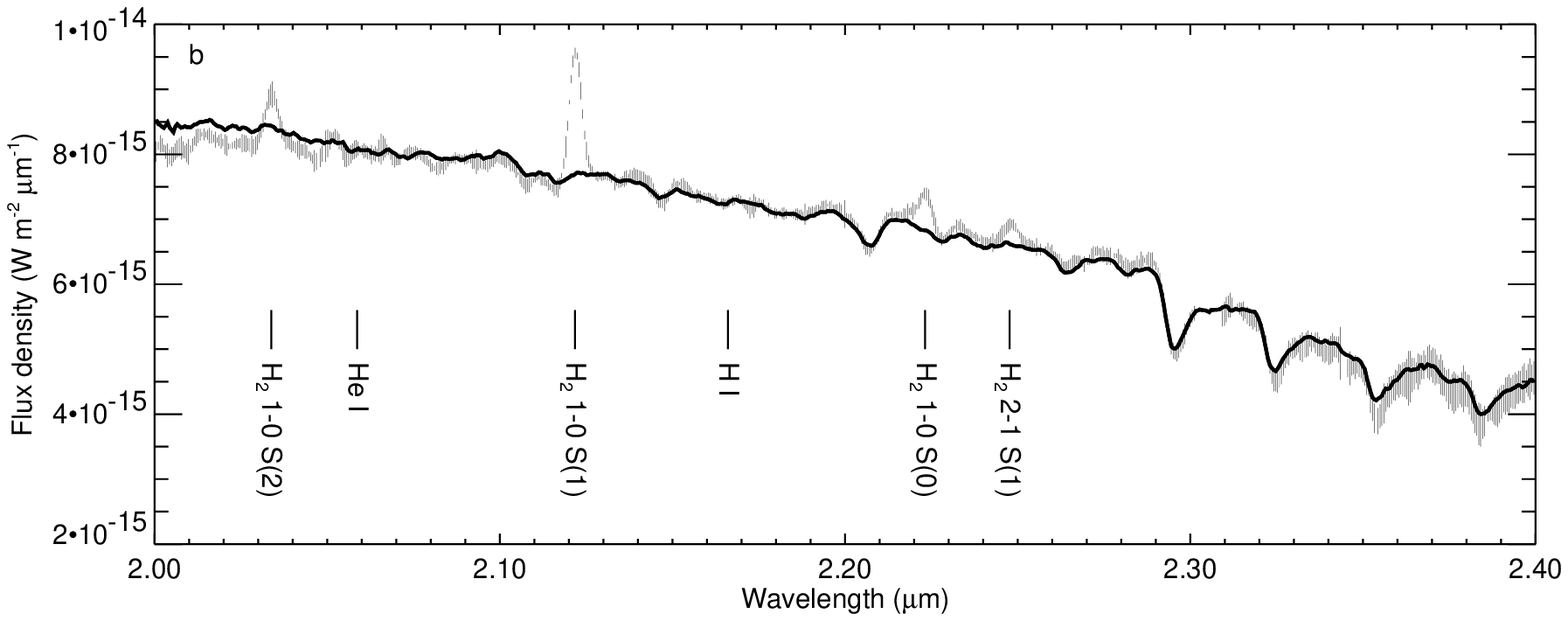}
\caption{The (a) H-band and (b) K-band spectra of NGC~5005.  A combination
of the composite quiescent spectrum and a power-law continuum have been 
overlaid as a solid line.  Note the strong shock excitation lines but the
absence of Brackett-$\gamma$ emission.}
\label{f_specn5005}
\end{figure}

\begin{figure}
\plotone{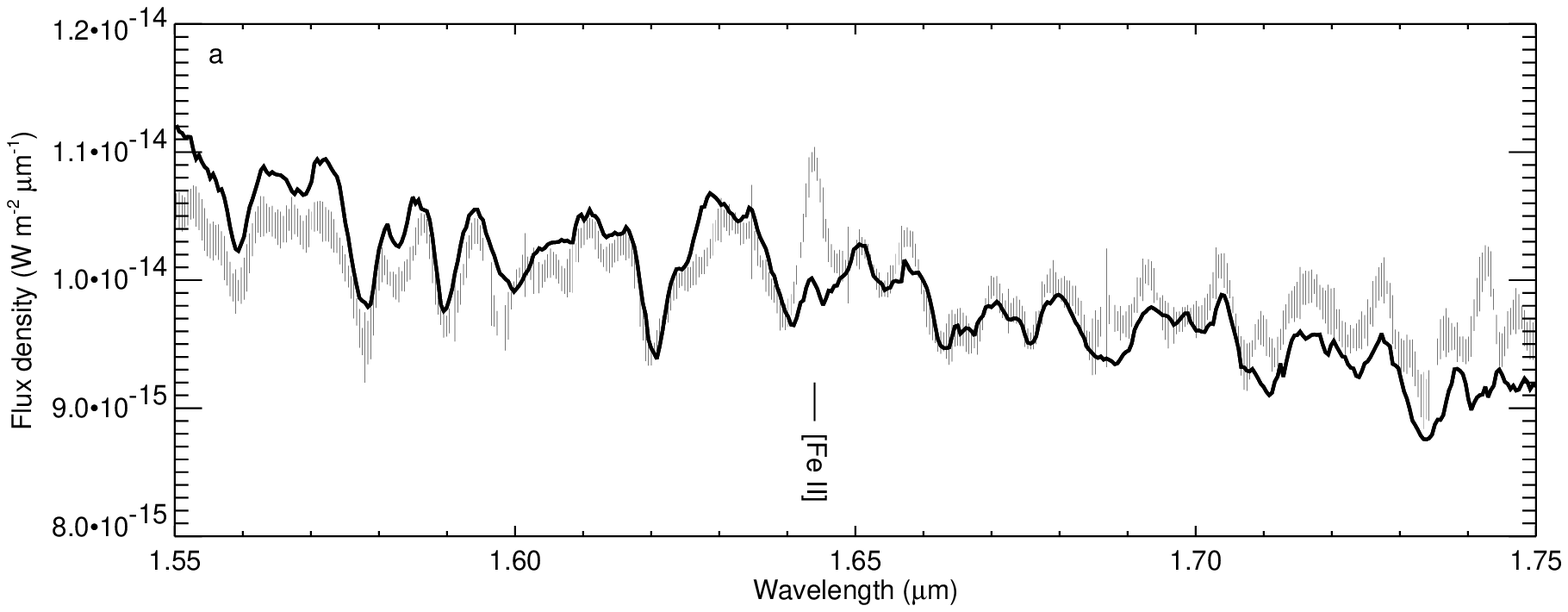}
\plotone{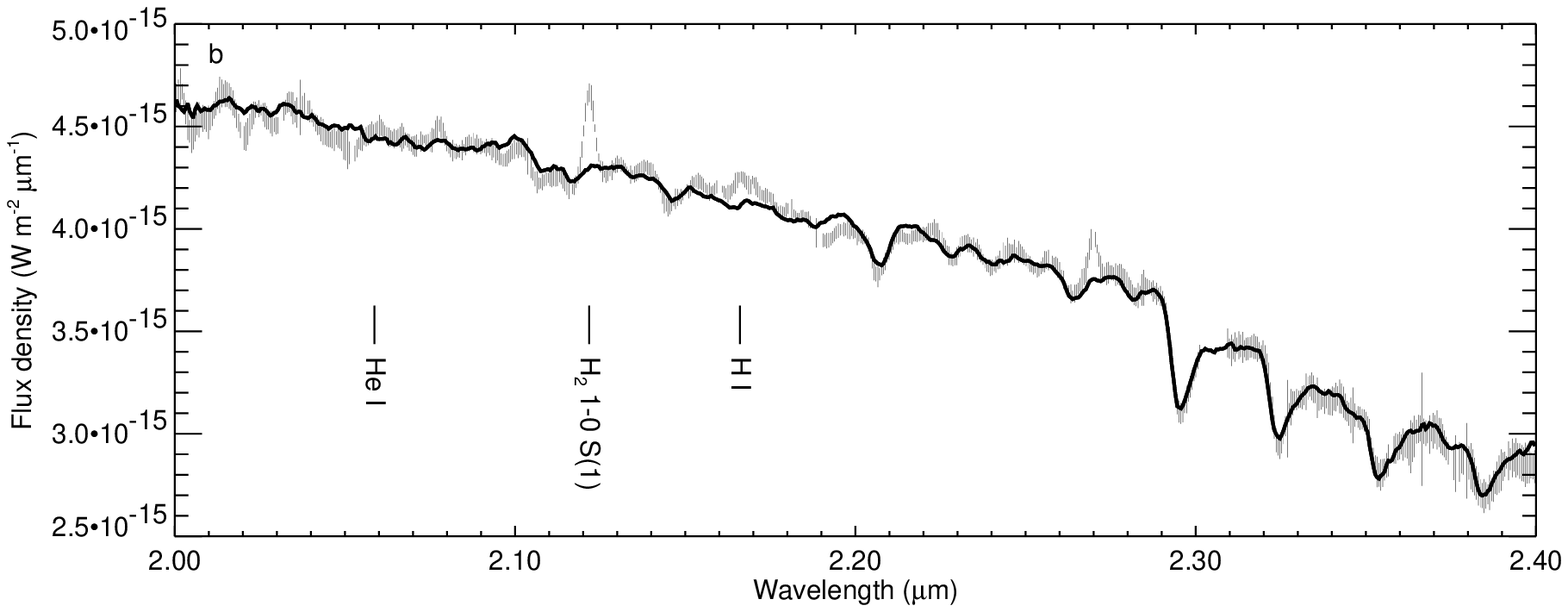}
\caption{The (a) H-band and (b) K-band spectra of NGC~5033.  As with NGC~5005,
a combination of the composite quiescent spectrum and a power law have been 
overlaid as a solid line in the K-band plot.  The strong shock excitation 
lines are the most obvious features in this spectrum, but a broad yet weak 
Brackett-$\gamma$ feature is present.}
\label{f_specn5033}
\end{figure}

Finally, the galaxies NGC~5005 and NGC~5033 exhibit strong
H$_2$ and Fe~{\small II} lines from shock excitation but either weak
or nonexistent Brackett-$\gamma$ emission.  Interestingly, the K-band
continua of these galaxies do not match the composite quiescent
spectrum.  The slopes of the two galaxies differ from the quiescent
spectrum, and the observed CO bands in NGC~5033 are shallower than
most other galaxies.  However, the continua can be fit by a linear
combination of the composite quiescent spectrum and a power law.
Table~\ref{t_n5005-5033pw} provides basic data on the underlying power
laws in the K-band.  Note that these power law fits are only
approximations as they are highly sensitive to the exact regions of
the K-band continuum used in the fits.  Figures~\ref{f_specn5005} and
\ref{f_specn5033} show the spectra for these galaxies along with the
best fitting continua.

The absence of strong recombination lines indicates that NGC~5005 and
NGC~5033 have at least evolved past 8~Myr.  If the shock excitation
lines are interpreted as originating from supernovae, then we could
conclude that the stellar populations in these galaxies are in the 8 -
36~Myr age range.  However, these two galaxies are interacting with
each other \citep{hst82}.  The shock excitation lines may simply be
tracing cloud collisions from interaction-induced gas infalls into
these galaxies' nuclei.  Since NGC~5005 is classified as a LINER and
NGC~5033 is classified as a Seyfert \citep{hfs97b} (HFS97b), another
possibility is that the shock excitation lines are the results of gas
outflows from AGN.  The presence of additional continuum emission
could be further proof of AGN in these galaxies.  However, when we
discuss the near- and mid-infrared colors and luminosities in
Section~\ref{s_specphotcomp}, we will discuss additional evidence that
the shock excitation lines may arise from supernovae.

\subsection{Extranuclear Targets and Complex Nuclear Targets 
\label{s_nonquies_complex}}

In this section, we discuss unusual targets with either complex nuclear
structure or unusual star clusters outside their nuclei that also had
spectra taken.  All of the galaxies listed below produce detectable
emission lines in at least one location.

\subsubsection{NGC 1569}

\begin{figure}
\plotone{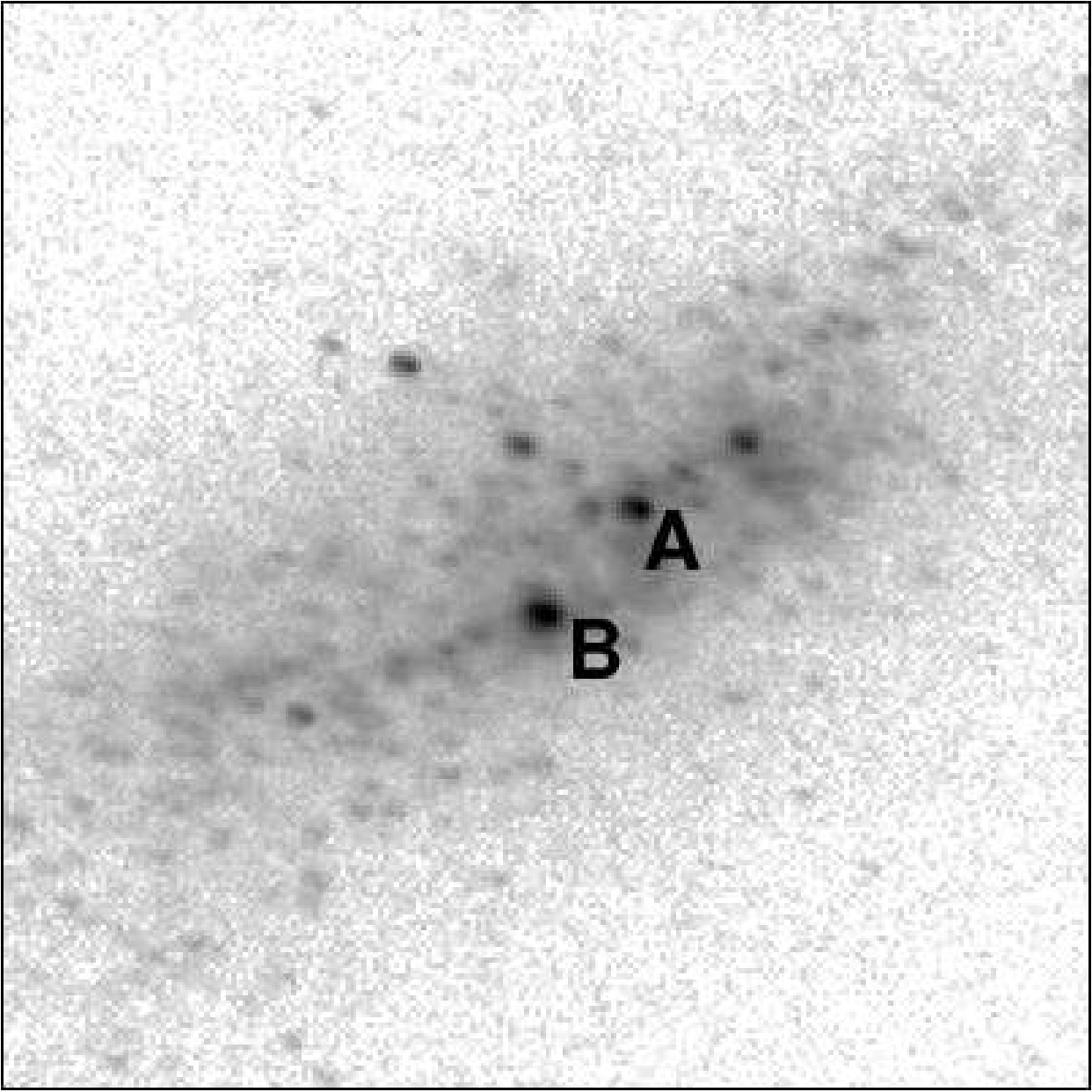}
\caption{A K-band image of the inner 60\arcsec of NGC~1569, with the A and 
B clusters identified in the image.  North is up and east is left.}
\label{f_imgn1569}
\end{figure}

NGC~1569 contains two prominent star clusters.  These star clusters
are identified in Figure~\ref{f_imgn1569}.  This is a well-studied
system where the stellar populations have been identified
\citep{gtcetal98, mhs01, aetal01, oetal01}, so these observations
provide few new insights into the galaxy itself.  However, the results
should be considered within the context of the sample as a whole,
particularly since this is one of the few galaxies in the sample where
emission lines have been detected in the spectroscopy.

\begin{figure}
\plotone{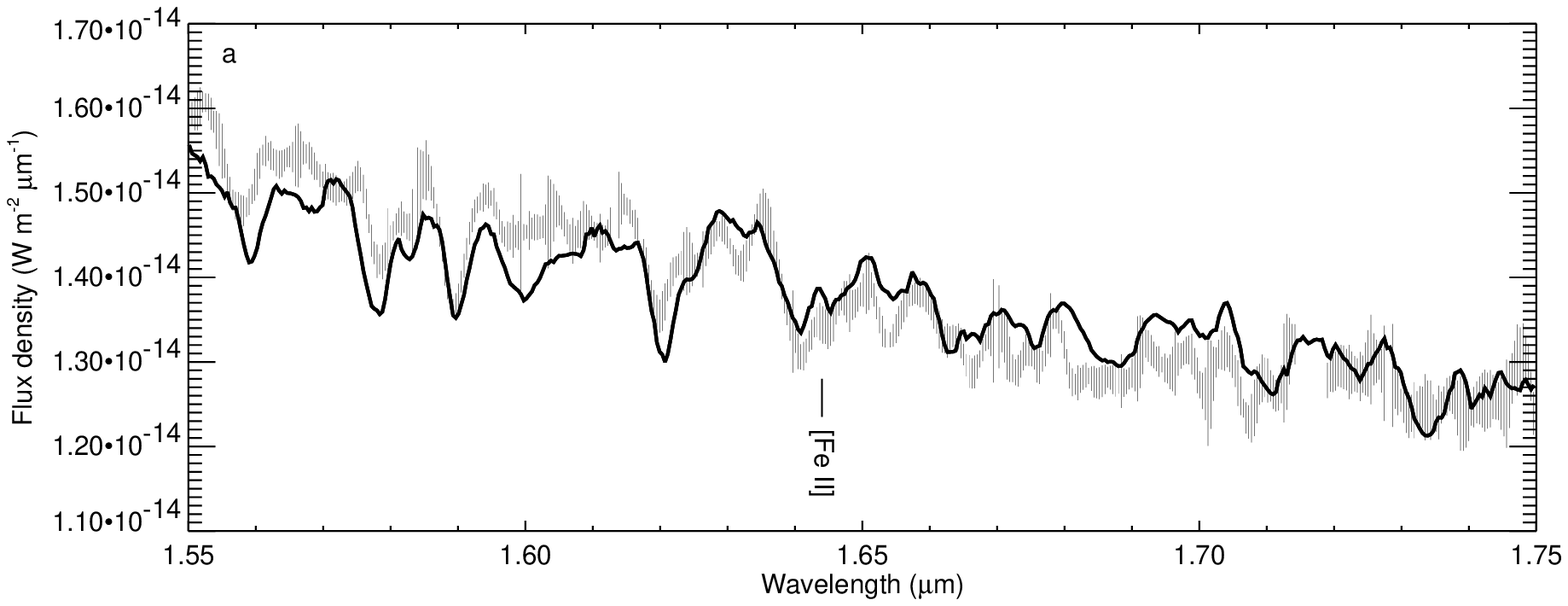}
\plotone{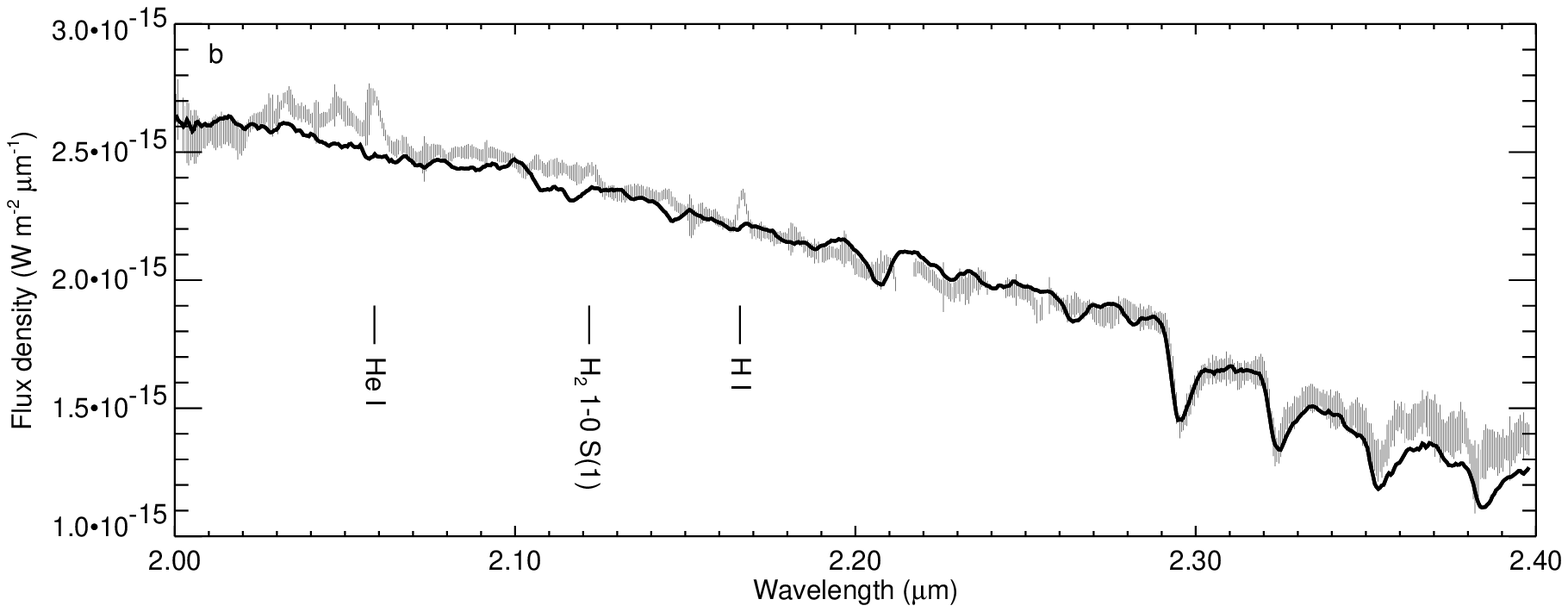}
\caption{The (a) H-band and (b) K-band spectra of cluster A in NGC~1569.
Note not only the presence of He~{\small I} and Brackett-$\gamma$ emission
lines but also the relatively weak metal lines in the K-band.}
\label{f_specn1569a}
\end{figure}

\begin{figure}
\plotone{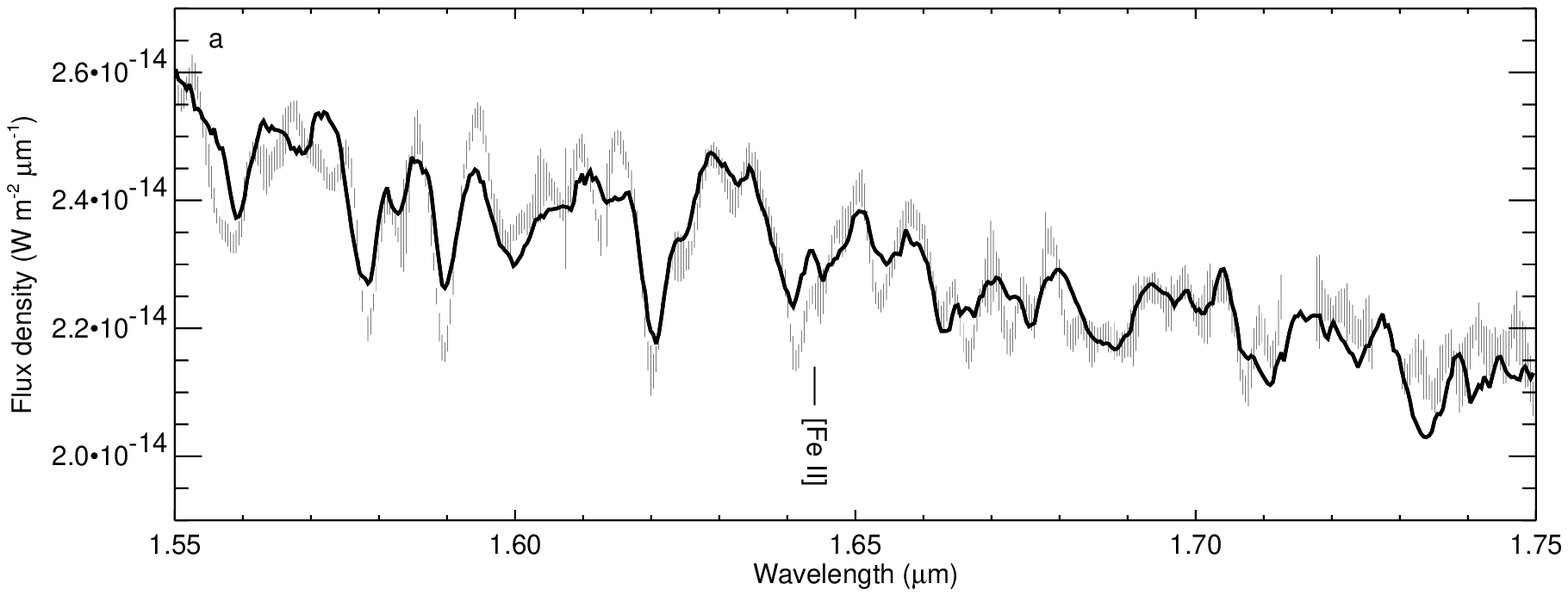}
\plotone{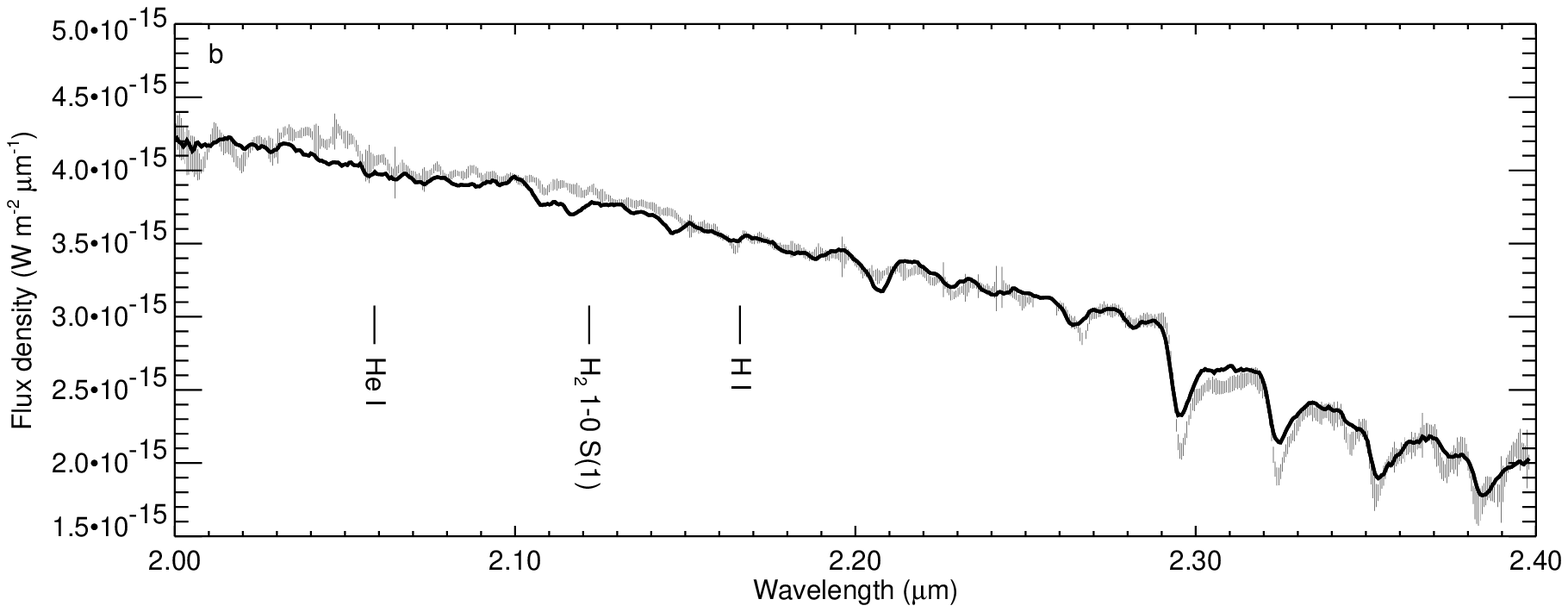}
\caption{The (a) H-band and (b) K-band spectra of cluster B in NGC~1569.
In contrast to the spectra for cluster A, these spectra lack the He~{\small I}
and Brackett-$\gamma$ emission lines.  However, they do include the same 
relatively shallow metal absorption features.  They also feature deeper CO
features than the overlaid composite quiescent spectrum, which indicates that
the continuum is dominated by red giants younger than 180~Myr.}
\label{f_specn1569b}
\end{figure}

Figures~\ref{f_specn1569a} and \ref{f_specn1569b} present the spectra
of NGC~1569A and NGC~1569B.  NGC~1569A exhibits strong
Brackett-$\gamma$ emission and He~{\small II} emission but no
detectable Fe~{\small II} or H$_2$ emission.  NGC~1569B, however,
produces no strong near-infrared emission lines. Otherwise, NGC~1569A,
along with NGC~1569B, have similar continua with similar absorption
features.

At first glance, the H and K continua appear to match the composite
quiescent spectrum very well.  However, a closer look at the spectral
line emission indicates that the absorption features in the NGC~1569
spectra do not match the composite quiescent spectrum.  Note, for
example, the missing Al~\small{I} and Mg~\small{I} features near
2.11~$\mu$m and the apparent absense of the Ca~\small{I} triplet near
2.26~$\mu$m.  These lines may all be missing because NGC~1569, as a
dwarf, has a lower metallicity than most nearby spiral galaxies.
Furthermore, note the deeper CO features in the spectrum of NGC~1569B.
This indicates that the B cluster may contain higher-mass red giants
than the average galactic nucleus and that it may be younger than
180~Myr.

Nonetheless, the data imply that a starburst less than 3.5~Myr old is
embedded within an older stellar population in NGC~1569A and that
NGC~1569B consists of an older stellar population.  These results are
consistent with most previous studies.  The lack of detection of shock
excitation lines in the near-infrared spectra is also consistent with
\citet{getal02}, which did not detect any supernovae at radio
wavelengths.

\subsubsection{NGC 3556}

\begin{figure}
\plotone{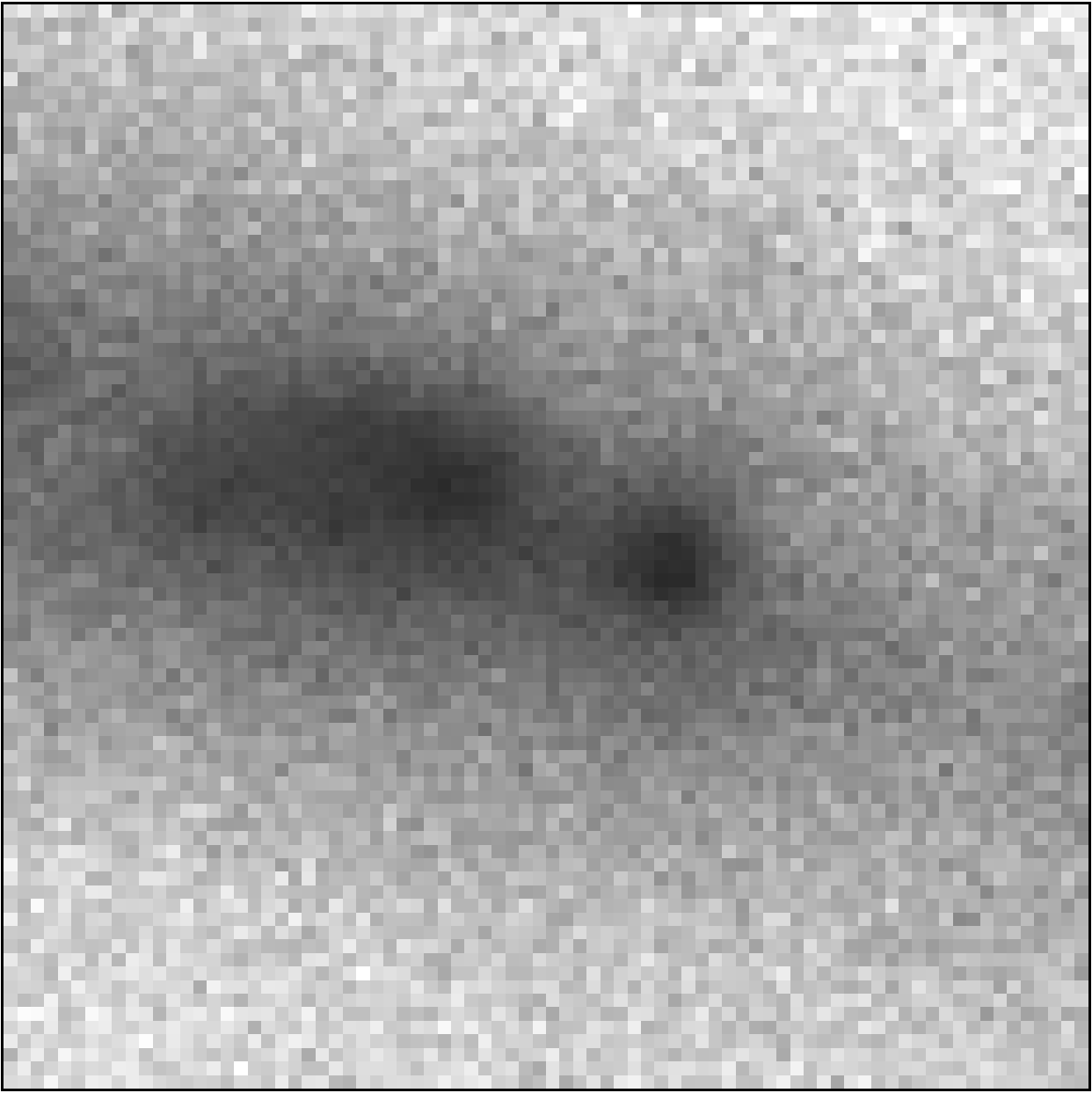}
\caption{A K-band image of the inner 15\arcsec of NGC~3556.  This corresponds
to a physical size of 1 kpc.  North is up and east is left.  
The east and west nuclear regions are both evident in this image.}
\label{f_imgn3556}
\end{figure}

The nuclear region of NGC~3556 contains two regions of similar surface
brightness but different shape.  Figure~\ref{f_imgn3556} shows a
K-band image of these regions, which we will refer to as the east and
west nuclear regions of NGC~3556.  Although the two regions are
relatively close to each other, their K-band spectra are stunningly
different.

\begin{figure}
\plotone{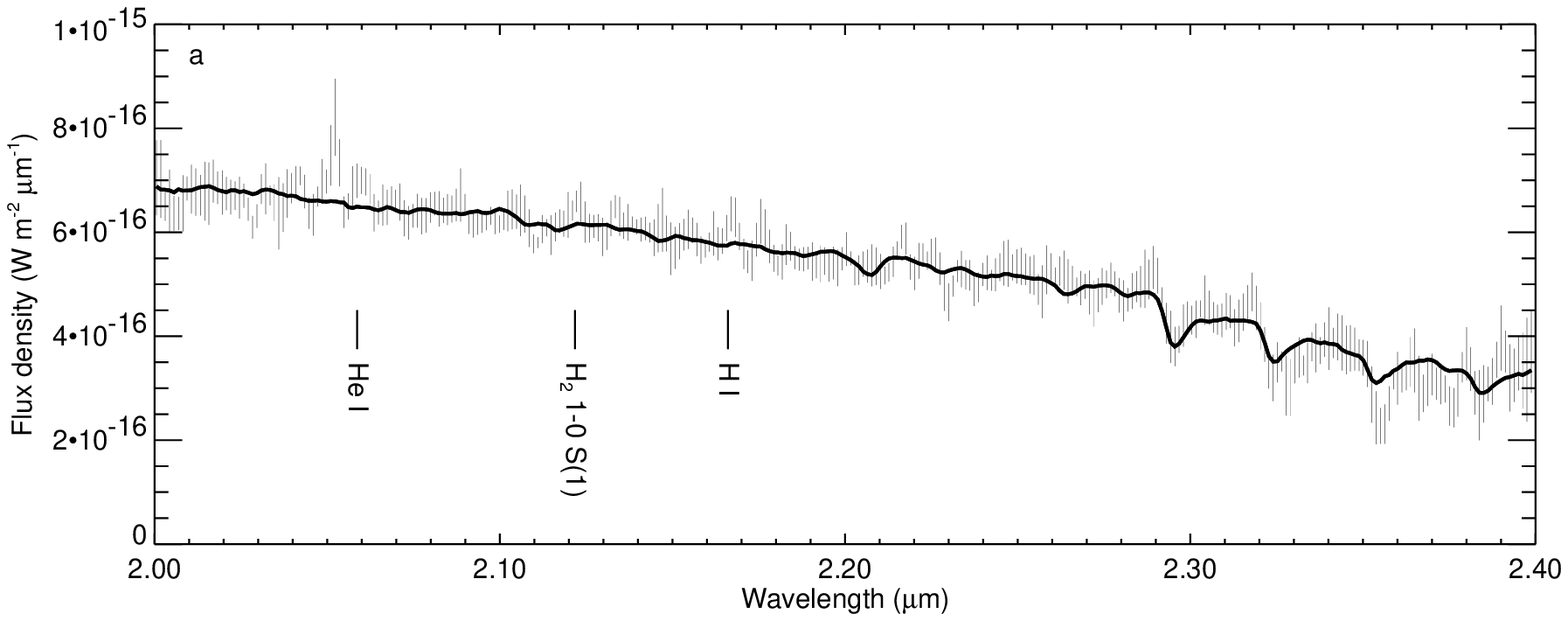}
\plotone{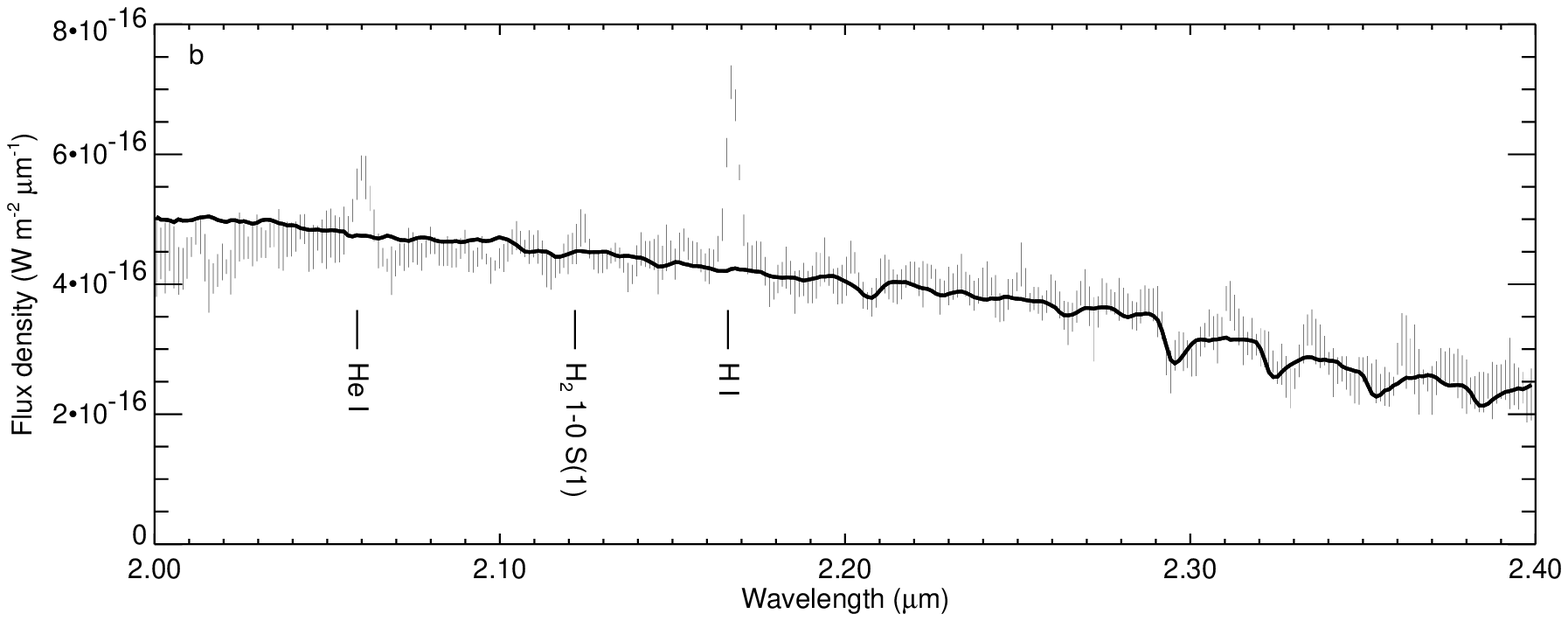}
\caption{The K-band spectra of (a) the east nuclear region and (b) the west
nuclear region of NGC~3556.  Brackett-$\gamma$ and He~{\small I} lines are
present in the spectrum for the west nuclear region but not the east nuclear
region.  (The narrow spike near 2.05~$\mu$m in the spectrum of the east nuclear
region appears to be artificial.)}
\label{f_specn3556}
\end{figure}

Figure~\ref{f_specn3556} shows the spectra of the east and west
nuclear regions, with the composite quiescent spectrum overlaid on
each spectrum.  The underlying continua for both objects has the same
slope as the composite quiescent spectrum and do feature similar CO
absorption features, although most other absorption features are so
poorly distinguished that the match between the spectra and the
composite is not as clear as in other cases.  The east nuclear region
appears to be a relatively quiescent region.  The west nuclear region,
however, exhibits strong He~{\small I} and Brackett-$\gamma$ line
emission that implies the presence of a stellar population that is
less than 3~Myr old.

\subsubsection{NGC 4100}

\begin{figure}
\plotone{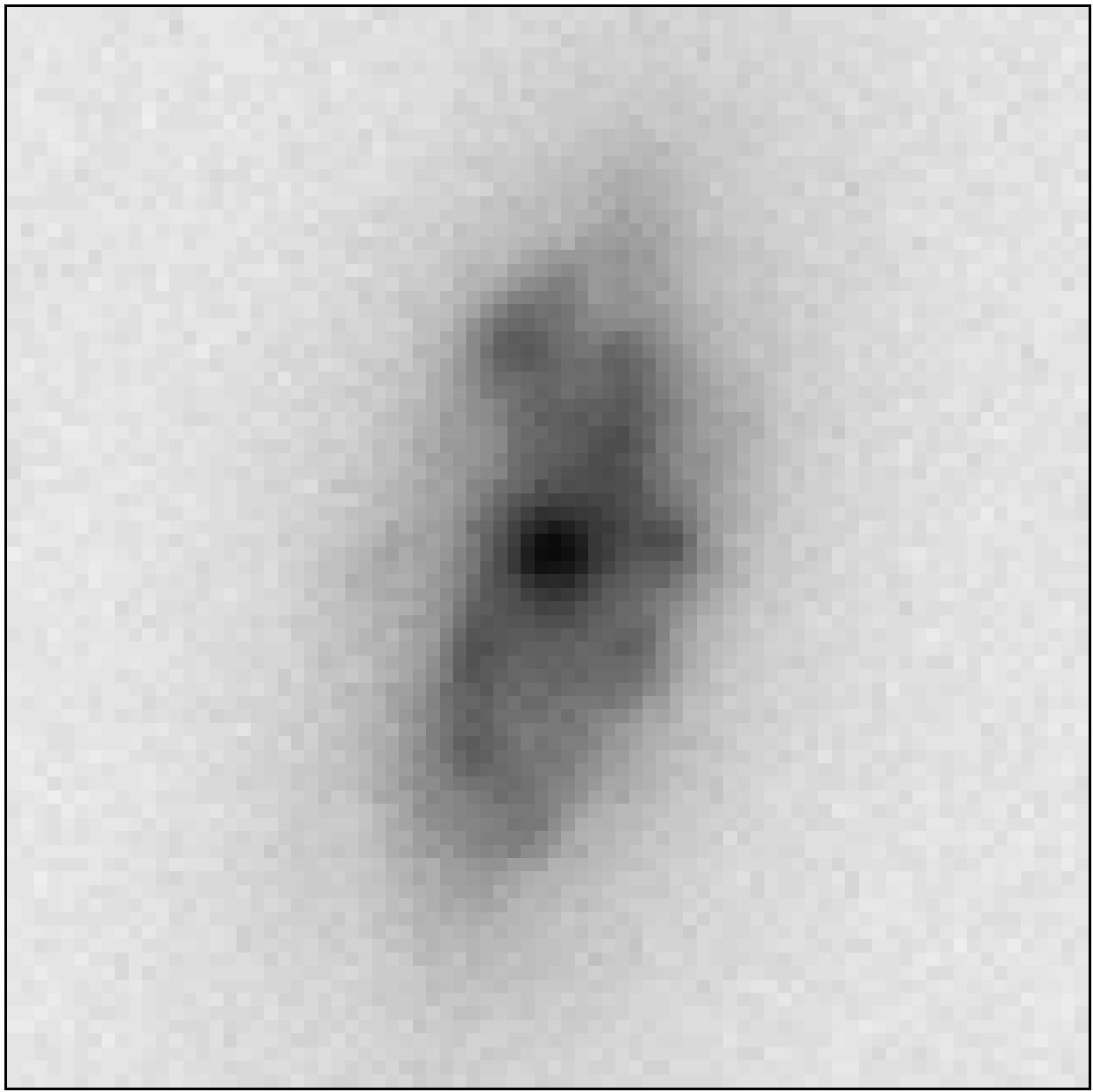}
\caption{A K-band image of the inner 15\arcsec of NGC~4100.  This corresponds
to a physical size of 1.2 kpc.  North is up and east is left.  Note the
knotted structure of this galaxy's nucleus.}
\label{f_imgn4100nuc}
\end{figure}

\begin{figure}
\plotone{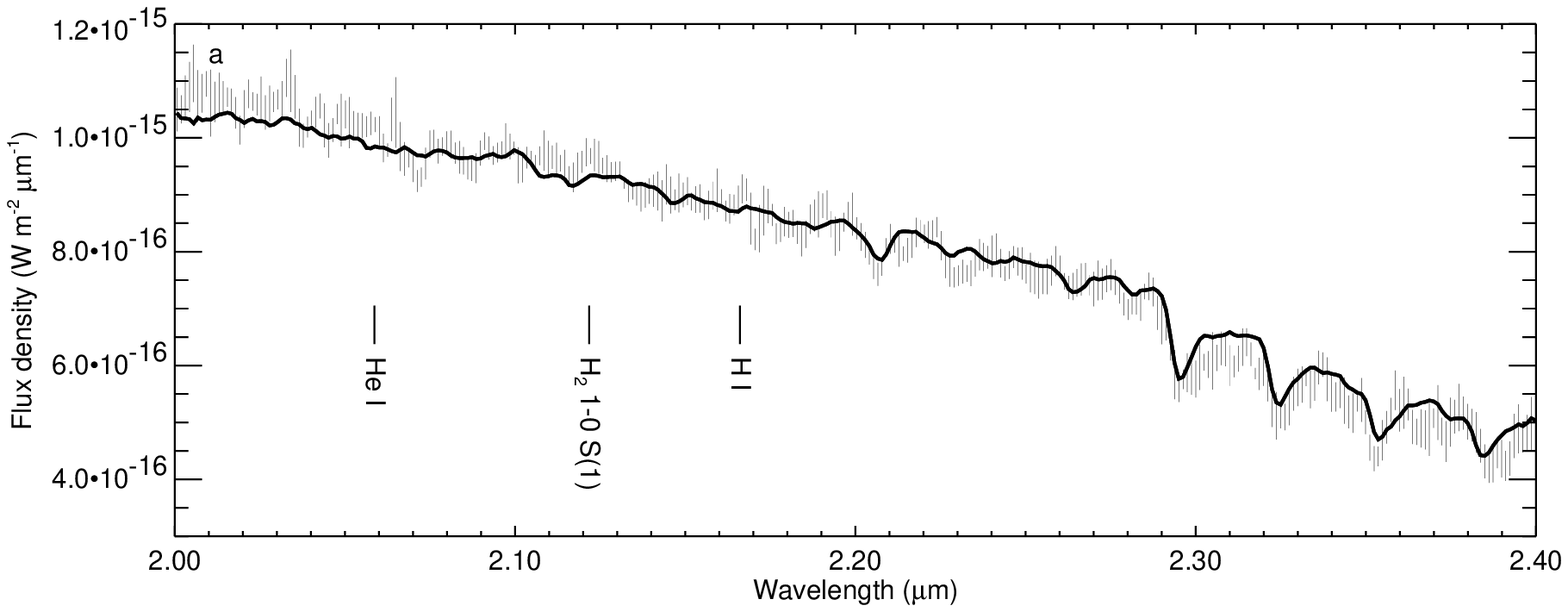}
\plotone{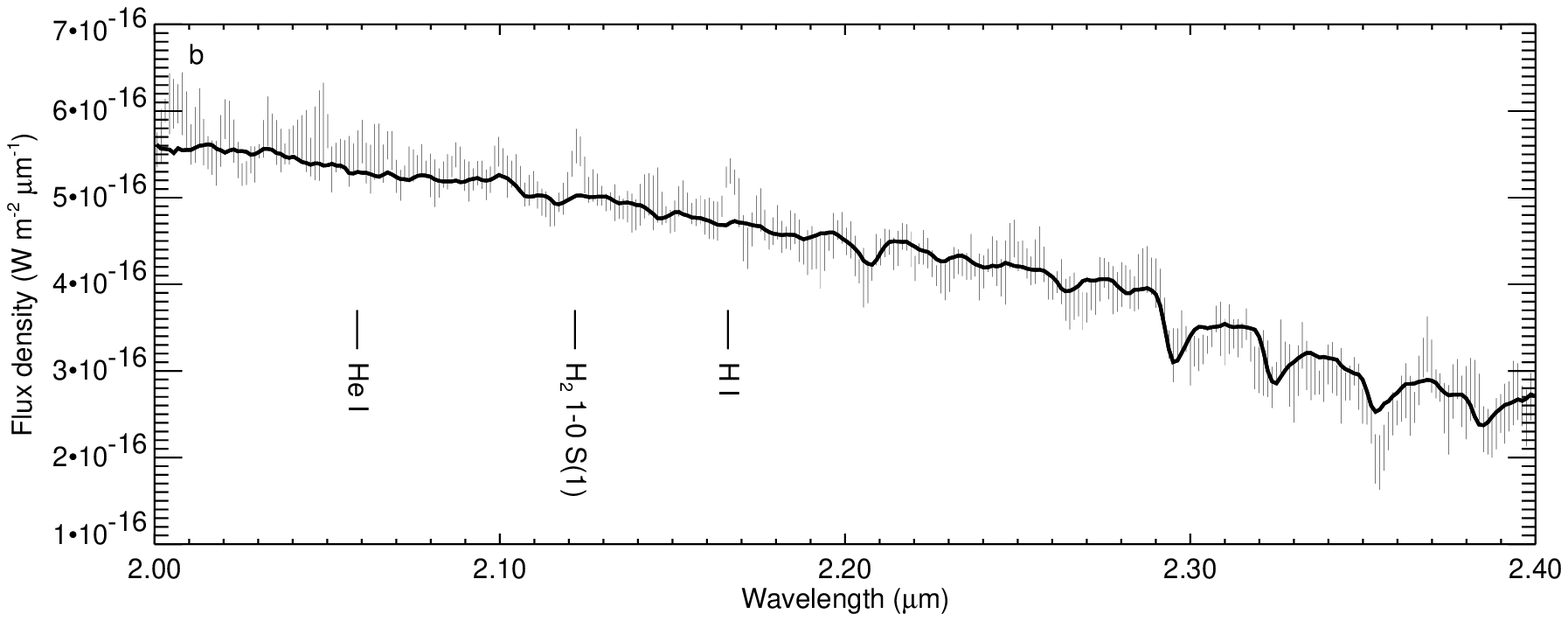}
\caption{The K-band spectra of (a) the nucleus and (b) the region 1.2\arcsec
south of the nucleus in NGC~4100.  The nucleus produces a quiescent spectrum,
but the region just to the south produces both photoionization and shock
excitation lines.}
\label{f_specn4100}
\end{figure}

Based on images and information from the literature, NGC~4100 seems to
be an unspectacular Sbc galaxy with neither Seyfert nor LINER nuclear
activity (HFS97b).  A close examination of the nucleus in the K-band,
however, reveals that the nucleus actually has a knotted structure, as
seen in Figure~\ref{f_imgn4100nuc}.  The spectroscopy of the target
also produces some intriguing results, as seen in
Figure~\ref{f_specn4100}.  The K-band spectrum of the center of the
galaxy simply appears quiescent.  The spectrum of the region just
south of the center of the galaxy, however, reveals the presence of
detectable Brackett-$\gamma$ and molecular hydrogen emission.  Gas
falling into the nucleus of this galaxy may have created the star
formation structure in this galaxy's center.  Given the complexity of
the structure, we suggest that the nucleus of NGC~4100 should be
observed with an integral field spectrometer so as to better
understand the locations of photionizing stars, evolved red stars, and
shocks in the system as well as to determine the overall masses in
young and old stars.  

The complex nuclear star formation region of NGC~4100 may be
significantly more extended than the slit used for the observations.
Therefore, we have declined to use NGC~4100 in any of the quantitative
analysis, and we do not report any emission line equivalent widths or
luminosities.  However, we still use NGC~4100 qualitatively in the
discussion.

\subsubsection{NGC 5676}

The observations of NGC 5676 not only allowed us to examine the
spectra of the nucleus but also of a bright knot of star formation
north of the nucleus.  This knot stood out prominently in the
12~$\mu$m image presented in Paper 1.  At that wavelength, the surface
brightness of the northern knot was higher than the surface brightness
of the nucleus.  This strongly contrasts with the K-band image of the
galaxy, where the nucleus clearly has a surface brightness that is
higher than any region within the disk even though the northern knot
is a prominent source of K-band emission in the disk.  Clearly, some
kind of extraordinary star formation event is taking place in the
northern knot.  The K-band spectroscopy reveals more clues as to
exactly what is happening.

\begin{figure}
\plotone{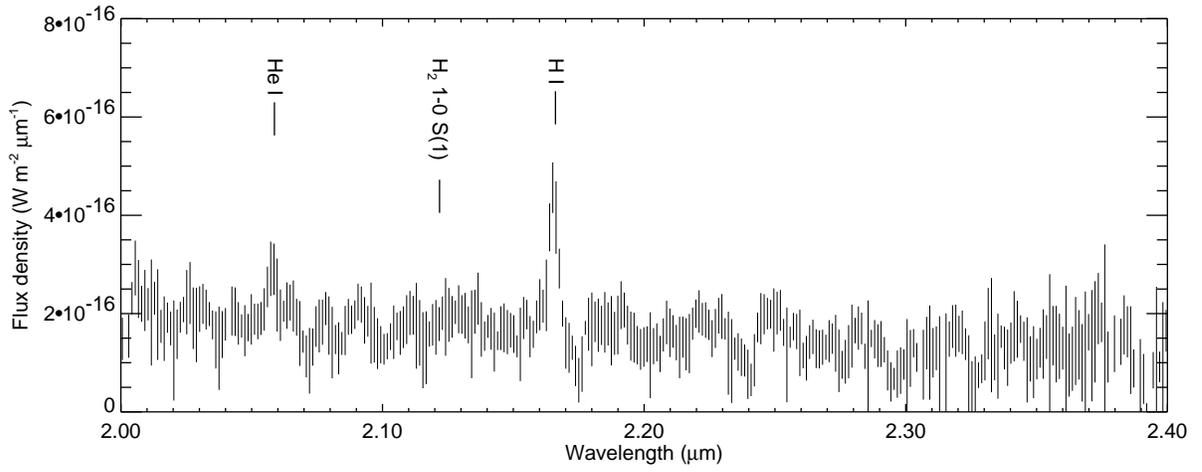}
\caption{The K-band spectrum of the northern knot of NGC~5676.  Note
the weak continuum and very strong Brackett-$\gamma$ and He~{\small I}
lines.}
\label{f_specn5676}
\end{figure}

Figure~\ref{f_specn5676} shows the K-band spectrum of the the northern
knot of NGC~5676.  The region produces very strong Brackett-$\gamma$
and He~{\small I} emission that dominates the K-band
emission. Otherwise, no other emission or absorption features are
discernable in the northern knot's spectrum.  The presence of the
He~{\small I} lines and the lack of H$_2$ shock excitation lines
suggests that this system is younger than 3~Myr.

Clearly, the northern knot is a large, young star formation complex
that may represent an important stage in the evolution of disk stars not
only in NGC~5676 but in other galaxies as well.  Further spectroscopic 
observations of the complex would allow for the
calculation of the total mass of this system and could even
be used to look for dynamical clues as to how this complex formed.

\subsection{Morphological, Environmental, and AGN Characteristics}

The occurence of H- and K-band emission lines in only a few galaxies leads us
to examine what factors lead to the production of these features.  We will
consider morphological and environmental factors that could possibly be 
triggers or conditions for the star formation activity we observed.  We will
also discuss the implications of our results for better understanding
AGN activity, particularly LINERs.

\subsubsection{Morphological Characteristics}

From looking at the morphological types of all the non-quiescent
galaxies, it is clear that location along the Hubble sequence is a
major factor.  Of the eight non-quiescent galaxies, all of them are
classified in RC3 as Sbc galaxies or later types.  This clearly
supports the paradigm that the relative strength of star formation
increases along the Hubble sequence, even though \citet{k_rc98} states
that this trend is primarily one seen in the disks of spiral galaxies,
not in their nuclei.  The finding agrees with the photometric results
reported in Paper 2.
This also agrees well with the results presented in \citep{mbpetal01},
where composite near-infrared spectra made for different morphological
types of galaxies show that recombination and shock excitation line
emission is most prominent in late-type galaxies.

The results presented here probably reflect the
change along the Hubble sequence in the relative contributions of bulge 
stars to the H- and K-band emission.  In S0s and early-type spiral galaxies, 
which are defined as having large bulges, the bulge stars dominate the 
near-infrared emission.  If any star formation was present, the recombination
line emission is relatively insignificant compared to the bulge star continuum
emission and therefore extremely difficult to detect.  In late-type spiral
galaxies, however, fewer bulge stars are present by definition.  Therefore,
if star formation is present, the lower continuum from the bulge stars allows
for the detection of recombination lines.

Nonetheless, not all late-type spiral nuclei produce detectable 
Brackett-$\gamma$ or shock excitation line emission, as quite a large number
of the near-infrared spectra of Sbc - Scd galaxies are quiescent.  
Although location along the Hubble sequence is a necessary factor for the 
H- and K-band line emission, it appears to not be a sufficient factor.

Bars have often been suggested as a means by which gas may be funneled into
the nuclei of galaxies, thus leading to enhanced star formation 
\citep{n88, a92, wh92, fb93, hs94, wh95}.
Several studies have found a difference between the nuclear star formation
activity of barred and unbarred spiral galaxies, although the differences are
mainly seen in early type galaxies \citep{d87, hetal96, hfs97a, retal01}.
Six of the eight non-quiescent galaxies in this sample are at least weakly 
barred according to RC3, although only three are identified as strongly barred.
However, two of the galaxies are identified as unbarred: NGC~4100 and NGC~5033.
The K-band image of NGC~4100 in Paper 1 hints at the possibility of a bar,
but NGC~5033 is unquestionably unbarred.  Given that many
of these non-quiescent galaxies are only weakly barred and that at least one
is unbarred, it seems as though bars are not absolutely necessary for 
triggering nuclear star formation.  Furthermore, some of the observed late-type
barred spiral galaxies have quiescent near-infrared spectra, suggesting that
bars do not directly correspond to enhanced star formation activity.

\subsubsection{Interaction History \label{s_nonquies_envir}}

Interactions have been proposed as a means of triggering or enhancing star
formation in the nuclei of galaxies by driving gas inward, even before the
interacting galaxies begin to merge \citep{mh96}.  Observations of nuclear
star formation confirm these effects \citep{ketal85, cm85, ketal87, wetal88}.

In the non-quiescent galaxies in this sample, four of them are interacting.
NGC~289 is interacting with a dwarf companion \citep{a81}, NGC~4088 is
interacting with NGC~4085 \citep{v83}, and NGC~5005 and NGC~5033 are 
interacting with each other \citep{hst82}.  The nuclear star formation 
activity in the three interacting barred galaxies could just as easily be 
explained as being caused by the interaction as its is explained as being 
caused by the bar, especially in light of such studies as \citet{n88} that
discuss bars forming in interactions and causing gas infall.  For NGC~5033,
which is unbarred, the interaction is the only obviious mechanism for 
explaining why signs of star formation (or at least shocked gas) are found
in the nucleus.  

What is particularly interesting is that NGC~5005 and NGC~5033 have
very similar spectra.  If the shock excitation line emission from
these two galaxies is interpreted as coming from supernovae produced
from starburst events, then the events may have occured
nearly-simultaneously in both galaxies.  This would be a powerful
demonstration of the effects of interactions on nuclear star formation
activity if it could be confirmed.

Two additional non-quiescent galaxies, NGC~1569 and NGC~5713, have
lopsided, amorphous appearances.  that suggest that they may have
experienced recent minor merger events.  Such minor merger events are
capable of enhancing nuclear star formation activity \citep{mh94,
hm95, rrk00}.  Therefore, we would suggest that these galaxies have
recently undergone minor mergers that has triggered their nuclear star
formation activity.

These results demonstrate that interactions are as important as bars
in triggering nuclear star formation activity.  NGC~3556 and NGC~4100,
however, are examples of non-interacting systems with current nuclear
star formation, which demonstrates that interactions are sufficient
but not necessary for triggering nuclear bursts.  Moreover, not all
interacting galaxies show near-infrared spectroscopic signs of recent
star formation activity.  Several of the quiescent galaxies in this
sample are also interacting, including NGC~1512 \citep{hetal79},
NGC~4725 \citep{h79}, NGC~5457 \citep{ddj80}, NGC~5566 \citep{hst82},
NGC~5746 \citep{sb94}, NGC~5846 \citep{sb94}, NGC~5866 \citep{sb94},
and NGC~5907 \citep{sb94}.  Many, but not all, of these quiescent
interacting galaxies are early type spirals with large bulges.
\citet{mh94} demonstrated that large bulges can inhibit the infall of
nuclear gas into galactic nuclei during interactions, which may
explain why most of the interacting quiescent galaxies in this sample
do not exhibit nuclear near-infrared line emission.

\subsubsection{AGN Characteristics \label{s_nonquies_agn}}

The relation between the AGN activity (as determined from optical spectroscopy)
and the H- and K-band spectroscopy can provide additional clues in 
understanding the underlying mechanisms behind the nuclear activity.  

In Seyferts, the question has been how the AGN activity may be related
to star formation.  Several studies, including ultraviolet and optical
observations \citep{gp93, hetal97, ghletal98, ghl01}, near-infrared
spectroscopic surveys \citep{oetal95, oetal99}, far-infrared surveys
\citep{rrj87, dmm88}, and CO surveys \citep{hetal89} have found
evidence for relatively recent circumnuclear star formation in Seyfert
2 galaxies.  Seyfert 1 galaxies show no signs of being associated with
star formation activity, although intermediate classes between 1 and 2
may be.  The emerging paradigm suggests that Seyfert 2 activity
appears on the order of 10~Myr after an instantaneous starburst.
According to Starburst99, supernova activity should be prevalent,
near-infrared recombination lines will be undetectable, and absorption
features will generally be deeper than average.  However, some
infrared surveys of Seyferts, such as \citet{oetal95},
\citet{oetal99}, and \citet{hetal99} note that the absorption features
could be diluted by continuum emission from the AGN.  Nonetheless,
these studies have presented us with a model that can be compared to
observations.

In LINERs, near-infrared spectroscopy is particularly important.  With
both photoionization and shock excitation lines, more constraints can
be placed on whether the optical spectra of LINERs are produced by
photoionization by either clusters of young stars \citep{tm85, s92} or
AGN \citep{hfs93} or by shock excitation from supernovae produced in a
starburst or AGN \citep{h80}.  As has been discussed by
\citet{aetal97}, near-infrared Fe~{\small II} and recombination lines
can be the key to distinguishing between shock excitation and
photoionization results.

We will rely on the classifications of HFS97b in this 
discussion.  Two of the eight non-quiescent galaxies in this sample are not 
classified by HFS97b, although the NASA/IPAC Extragalactic Database lists one
of them as an H~{\small II} galaxy.  Four of the six non-quiescent galaxies are
also identified by HFS97b as H~{\small II} galaxies.  This leaves one Seyfert
(NGC~5033) and one LINER (NGC~5005) as AGN with detected H and K spectral 
line emission.

Only two galaxies observed in this spectroscopic survey were
classified as Seyferts by HFS97b.  The near-infrared spectrum of
NGC~4725 was quiescent.  HFS97b, however, state that the
classification of NGC~4725 is uncertain; perhaps the non-detection of
emission lines along with the lack of a continuum that dilutes the
absorption features indicates that this galaxy is not a Seyfert.  The
near-infrared spectrum of NGC~5033, however, exhibited weak but broad
Brackett-$\gamma$ emission as well as strong shock excitation line
emission.  The broad Brackett-$\gamma$ line clearly indicates the
presence of an AGN.  The shock excitation lines, however, could be
interpreted as originating from the AGN, from the interaction with
NGC~5005, or from a circumnuclear starburst.  If it is a circumnuclear
starburst, this would show the same kind of association between
Seyfert optical spectral activity and star formation activity that had
been found previously.  Unfortunately, this sample lacks enough
Seyferts to really make a definitive statement about them, although
many other surveys in the literature (see references in
Section~\ref{s_intro}) could make stronger statements.

Seven of the galaxies in this spectroscopic sample were classified as
LINERs by HFS97b.  Another nine were classified as transition objects,
which may be more closely related to LINERs than to other classes of
objects. However, near-infrared emission lines were only detected in
one of these 16 objects, NGC~5005.  This is a relatively low rate of
detection compared to the H~{\small II} galaxies, where 4 of 13
galaxies in the sample classified as such by HFS97b had detected
near-infrared line emission.  However, these results are consistent
with other studies; \citet{letal98} and \citet{stl01} also present
near-infrared LINER spectra that have no detectable emission lines.

In NGC~5005, the characteristic strong shock excitation lines,
particularly the Fe~{\small II} lines, and the absense of
recombination lines are consistent with other near-infrared
spectroscopy observations of LINERs \citep{letal98, aetal00}.  In this
target, it is obvious that shocks are propagating through the nucleus,
photoionization is relatively weak, and therefore the optical LINER
emission is probably generated by shock excitation.

In the other 13 objects, though, the absence of all near-infrared
emission lines and the close match between the near-infrared spectra
of these galaxies and the near-infrared spectra of other quiescent
galaxies indicate that the spectra are dominated by evolved red stars.
If shock excitation or photoionization by hot young stars is taking
place in the nuclei of these LINERs, the emission is weak compared to
the continuum emission from the older quiescent stellar populations.
If the non-detection of near-infrared spectral lines are interpreted
as the absence of shocks or star formation, then only weak AGN may be
responsible for the optical emission line spectra.

The relatively low nuclear 12~$\mu$m flux to K-band flux ratios
determined for the galaxies in this sample in Paper 2 also indicated
that nuclear star formation is relatively weak in LINERs and
transition objects compared to H~{\small II} galaxies.  These
additional spectroscopic observations demonstrate that evolved red
stars dominate the near-infrared emission in these systems and that
photoionization from star formation is unlikely to be causing the
optical LINER emission observed in these galaxies.

\subsection{Ratio of Young to Old Stellar Mass}

\subsubsection{Determination of the Young Stellar Mass}

The total mass formed within the most recent starburst event can be
determined using measurements of the Brackett-$\gamma$ or Fe~{\small
II} 1.644~$\mu$m luminosities.  We will use data from Starburst99 to
determine the conversion factors from these line luminosities to young
stellar masses.  We treat the star formation as though it takes place
in an instantaneous burst.  We assume that the metallicities are solar
and that the IMF of the burst has a slope of 2.35 and an upper mass
cutoff of 100~M$_\odot$.  Note that, for convenience, the data used to
calculate the conversion values correspond to a simulated burst that
forms $10^6$~M$_\odot$ in stars.  The luminosities and masses
presented below are dependent on the initial mass, but the final
conversion factors are mass-independent.

We will begin by finding conversion factors between Brackett-$\gamma$
luminosity and total young stellar mass.  As we stated above in
discussing the ages of the young stellar populations, the appearance
or nondetection of certain spectral lines indicate different ages for
the systems.  We therefore divide the galaxies with Brackett-$\gamma$
line detections into two categories: pre-supernova systems (systems
with ages of 0.0 - 3.5~Myr that exhibit no shock excitation lines) and
supernova systems with recombination line emission (systems with ages
of 3.5 - 8.0~Myr that have detectable shock excitation lines).
Table~\ref{t_brgconvert} presents Starburst99 data on the range and
median value of Brackett-$\gamma$ luminosities, the stellar masses
remaining in the systems, and the conversion factors from
Brackett-$\gamma$ luminosities to masses for these two types of
systems.  Note that the large change in the Brackett-$\gamma$
luminosities in the 3.5 - 8.0~Myr range mean that the young stellar
masses determined from the line luminosities will only be accurate to
within a factor of 10.

To use the Fe~{\small II} 1.644~$\mu$m emission lines to trace young
stellar masses, we must select a conversion factor that changes
supernova rates into Fe~{\small II} line luminosities.  We will use
$1.2 \times 10^{34}$~W~yr, which was determined by \citet{aetal03}.
This agrees within a factor of 2 with the conversion factor of $1.9
\times 10^{34}$~W~yr determined in \citet{vetal93} (using their
assumptions on the scalar factor $\nu$ and assuming that their
(E$_0$/$10^{51}$ erg) term is equal to unity).  These conversion
factors, however, disagree with the conversion factor of $1.7 \times
10^{33}$~W~yr determined in \citet{c93} (using their assumptions
presented about the Fe~{\small II} luminosity of supernova remnants
and the duration of Fe~{\small II} emission).  However,
\citet{aetal03} and references therein state that the Fe~{\small II}
emission should last longer than what is assumed in \citet{c93}, which
could partly explain why the \citet{c93} conversion factors are
inconsistent with other values.

As with the galaxies with Brackett-$\gamma$ emission, galaxies with
Fe~{\small II} emission can be split into two groups: supernova
systems with recombination line emission (which have ages of 3.5 -
8.0~Myr) and supernova systems without recombination line emission
(which have ages of 8.0 - 36~Myr).  Table~\ref{t_feiiconvert} presents
data on the range and median of supernova rates, the corresponding
median Fe~{\small II} 1.644~$\mu$m line luminosities, the total
stellar mass remaining in the systems, and the line luminosity-to-mass
conversion factors.

To calculate the ratios of young to old stars in the non-quiescent
galaxies using the K-band flux to trace the older stars, we will need
to subtract the total contribution of the young stars.  The K-band
flux within younger stars can simply be determined by multiplying the
Brackett-$\gamma$ and Fe~{\small II} line fluxes by ratios of the
K-band continuum fluxes to line fluxes predicted by Starburst99.
Table~\ref{t_contlinerat} presents these data for the three different
age ranges discussed above.  Note that these correction factors are
only good to within a factor of 10, mainly because the K continuum
flux varies by a factor of 10 within the selected time periods.
However, as we will see below, these values will work for the science
that we are doing.

\subsubsection{Determination of the Young to Old Stellar Mass Ratios}

Using the K continuum luminosity to old stellar mass conversion factor
derived in Section~\ref{s_quies_synth} and these conversion factors
between line luminosities and young stellar mass, we can now measure
both the mass produced in these nuclear star formation events (for at
least the regions falling within the slit) and the ratio of young
stellar mass to total stellar mass in the nuclei of these galaxies.
For objects such as Cluster A in NGC~1569, the west nuclear region in
NGC~3556, and the northern knot in NGC~5676, we will treat the
underlying continuum as though it originates from an evolved stellar
population of solar metallicity for simplicity, even though this may
not be entirely accurate.

Table~\ref{t_youngoldrat} lists the masses of young and old stars
within the non-quiescent systems and the ratio of the masses as well
as all of the parameters needed to calculate the masses.  Note that
the luminosities and masses in Table~\ref{t_youngoldrat} are for the
region that falls within the spectrometers' slits and may not
necessarily reflect the true total luminosities and masses for the
nuclear star forming regions if the regions are extended.  The results
demonstrate that a strong star formation event could produce up to
5~\% of the stellar mass found within a galaxy's nucleus.  The data
also demonstrate that we were able to detect star formation within
galaxies where the young stars accounted for less than 0.03~\% of the
total mass, but typically young stars make up $\sim$2~\% of the
nuclear stellar mass in a non-quiescent galaxy.  The young stars also
contribute appreciably to the K-band luminosities of these galaxies'
nuclei.  In some cases, the young stars contribute more than 10~\% of
the flux in the K-band.

These ratios of young to old stellar mass agree with an overall
picutre of nuclear stellar populations being formed in a series of 
instantaneous bursts.  Consider that 20~\% of the galaxies in this sample 
have recently increased their central masses by 2~\%.  These young stellar 
masses formed within the past 36~Myr.  This means that the mass of the average 
galaxy's nuclear population increases by $\sim$0.4~\%
over a period of 36~Myr.  The age of the universe is $\sim$14~Gyr.
Therefore, if the nuclear stellar masses are being increased continuously
over time, then 36~Myr~/~14~Gyr of the mass, or 0.25~\%, of the stellar mass
should have been created within the past 36~Myr in the average galaxy.  Despite
the number of approximations in this calculation, these two percentages are
consistent with each other, which supports the paradigm of nuclear star
formation occurring as instantaneous bursts spread over the lifetime of the
galaxy.

The data also place the northern knot of NGC~5676 into an interesting context.
The mass of stars being formed within the northern knot and the percentage
of mass in new stars is comperable to those values for nuclear bursts of 
star formation.  Although nuclear star formation is very important in nearby
galaxies, this star formation event in NGC~5676 highlights the importance
of not ignoring the role of extranuclear bursts of star formation in
galactic evolution.

As a more technical achievement, the analysis for both the Brackett-$\gamma$
and Fe~{\small II} lines produce similar results in the two galaxies in this
analysis.  For NGC~289, the two young stellar masses calculated from the
two lines are within a factor of 3 of each other.  For NGC~5713, however, the
results agree to within 20~\%.  This demonstrates that our methods for 
calculating young stellar masses are at least internally consistent.
However, it is quite a spectacular achievement, especially given some of the
order of magnitude uncertainties in the interpretation of the Starburst99
data.

\section{Comparison of Quiescent and Non-Quiescent Galaxies in 
Mid-Infrared / K Color-Luminosity Plots \label{s_specphotcomp}}

In Paper 2, we justified using mid- and far-infrared fluxes divided by
near-infrared fluxes as a tracer of star formation activity by examining the
ratios that were predicted by Starburst99.  As a test of these tracers,
we will compare the colors and luminosities of the quiescent and non-quiescent
galaxies to each other.

\begin{figure}
\plottwo{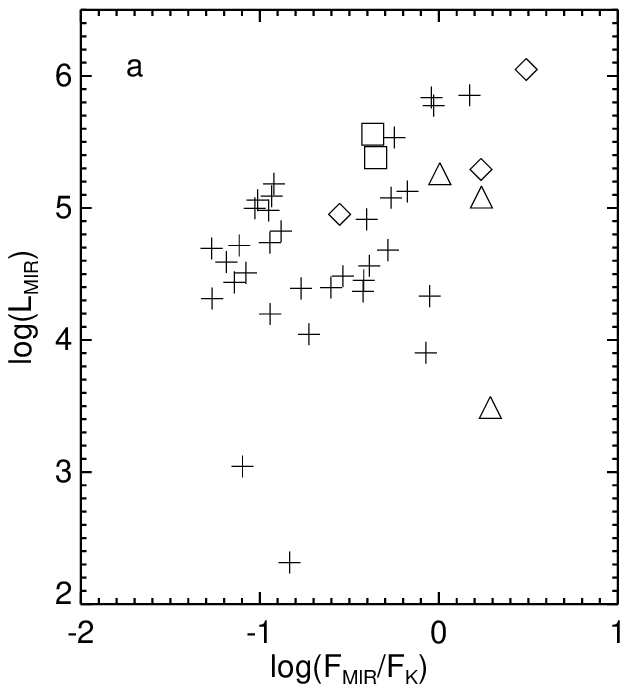}{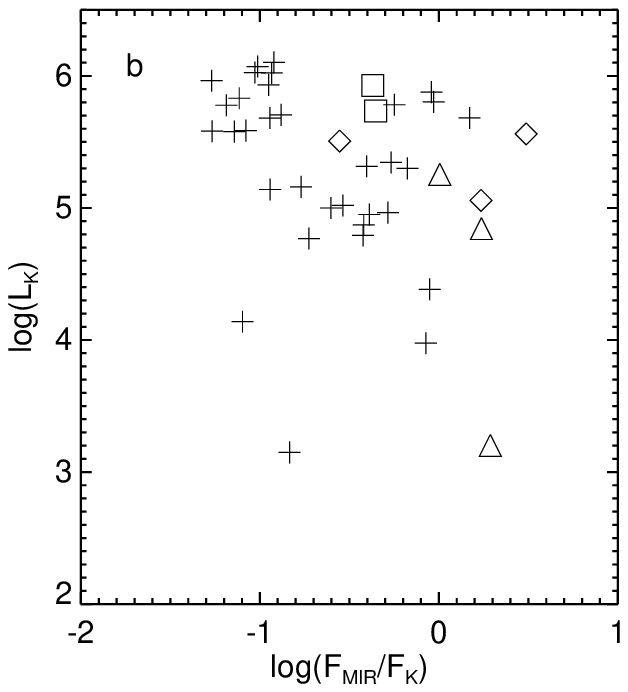}
\caption{The (a) mid-infrared luminosities and (b) K-band luminosities 
plotted against ratio of the mid-infrared to K-band fluxes.  The luminosities
are in units of L$_\odot$.  The crosses represent quiescent objects, while
the open symbols represent non-quiescent objects.  The triangles indicate
non-quiescent galaxies with only recombination line emission.  The diamonds
indicate galaxies with recombination and shock excitation line emission.
The squares indicate galaxies with only shock excitation line emission.}
\label{f_collum}
\end{figure}

Figure~\ref{f_collum} shows a plot of the $\frac{f_{MIR}}{f_K}$ colors
for the inner 15\arcsec of the galaxies in the spectroscopic subsample
against both the mid-infrared and the K-band luminosities from the
same regions.  Galaxies with quiescent and non-quiescent nuclear
spectra are represented as different symbols on the plots.  The
detection of spectral line emission is not correlated with luminosity
in either waveband.  However, when the non-quiescent systems with only
shock emission lines are excluded, it appears that ratios like the
$\frac{f_{MIR}}{f_K}$ ratio are well correlated with the detection of
recombination lines in the near-infrared, although the correlation is
not perfect.  When the K-S test is applied in comparing the quiescent
versus non-quiescent $\frac{f_{MIR}}{f_K}$ ratios, the resulting
fractional probability that the two data sets came from the same
samples is 0.026.  However, when the galaxies producing only shock
excitation line emission are left out of the non-quiescent sample, the
K-S test determines the probability as 0.0032.  This is a strong
statistical demonstration that shows the relation between systems
producing strong Brackett-$\gamma$ emission and systems with high
$\frac{f_{MIR}}{f_K}$ ratios.

Three of the non-quiescent galaxies, however, occupy locations in these
color-magnitude diagrams where their colors are close to the mean value but
where their luminosities are relatively high compared to galaxies with similar
colors.  All three of these galaxies have strong shock excitation line 
emission, which could imply that supernovae are present.  Now, consider
Figure~2 in Paper~2.  For an instantaneous burst, any tracer of the
bolometric luminosity (such as the 12~$\mu$m emission, which is discussed in 
Paper 2) when normalized by the K-band emission will drop sharply when
massive main sequence objects evolve into red supergiants.  The low colors are
found at $\sim$8~Myr, when supernova activity is also strong.  Therefore,
supernovae could be responsible for producing the observed shock emission lines
in these systems as well as reduced $\frac{f_{MIR}}{f_K}$ ratios and the
relatively high luminosities observed in these galaxies.

In summary, the $\frac{f_{MIR}}{f_K}$ ratio appears to be well correlated with
strong photoionization emission in the near-infrared, 
although it is not correlated with strong shock excitation emission.
Therefore, mir-infrared fluxes, as well as the correlated far-infrared fluxes,
can be used to trace star formation activity when normalized by K-band fluxes.

\section{Conclusions}

\subsection{Summary of Results}

In this sample, 33 of the 41 galaxies nuclei were found to have
quiescent spectra.  These spectra consisted of continuum emission with
many metal and CO absorption lines, which are characteristic of
evolved red stars.  Using Starburst99, we found that the composite
spectrum made from these quiescent galaxies' spectra was best
represented by an instantaneous burst that took place $\sim$180~Myr
ago, although we caution that this is not necessarily supposed to
reflect the true star formation history.  We found few differences
between using Salpeter and other IMFs, but we did find that the
continuous star formation scenario failed to reproduce the observed
equivalent widths.

The remaining 8 galaxies have nuclear spectra that include emission lines.
In many cases, the underlying continuum exhibited the same slope and 
contained the same absorption features that were present in the composite
quiescent spectra.  On top of this quiescent continuum, the galaxy may produce
either photoionization lines such as Brackett-$\gamma$ and He~{\small I}
emission, shock excitation lines such as H$_2$ lines and Fe~{\small II} 
emission, or both.  The lines present would reflect the age of the young
stellar population, with photoionization lines present up to 8~Myr after
a burst and shock excitation lines present from 3.5 to 36~Myr after a burst.
Some of the systems showed relatively point-like spectral line emission, 
whereas others exhibited emission in complex structures.  One galaxy, where
the nuclear emission was quiescent, had a strong extranuclear starburst that
was detected and included in some parts of the analysis.

These non-quiescent galaxies share many characteristics.  All of them are
Sbc galaxies or later, which suggests that these instantaneous nuclear
bursts are more likely in the later-type galaxies.  6 of the 8 galaxies were
definitely barred, which suggested that bars may play a role in triggering 
nuclear star formation.  However, 6 of the 8 galaxies either were interacting 
or had been involved in interactions, which suggests that interactions are just
as important as bars in triggering nuclear star formation.  In relation to
AGN activity, it appears that most non-quiescent galaxies are H~{\small II}
galaxies and that most LINERs and transition objects are quiescent.

The average ratio of young to old stellar masses for the non-quiescent
systems was found to be $\sim$2~\%.  Given that these 2~\% enhancements occur
in $\sim$20~\% of the galaxies in the sample, these enhancements are found
to be consistent with nuclear stellar populations forming through a series
of instantaneous bursts.

Finally, we found that the $\frac{f_{MIR}}{f_K}$ ratios functioned very well
in separating non-quiescent galaxies from quiescent galaxies.  The exception
was with galaxies that contained only shock excitation lines, but the
ratio was expected to decrease signficantly for these galaxies.

The data and results presented here now form a baseline for comparison to
other galaxies with more exotic star formation or AGN activity.  Furthermore,
the composite quiescent spectra and the conversion factors developed here
should provide additional analysis tools for studying enhanced star formation
activity.

\subsection{Future Work}

While the spectroscopy presented here surveys a range of different kinds
of galaxies, it is only a limited sample.  A broad range of morphologies
were included in the sample, but not a broad range of AGN types; the
spectroscopic sample only contained two Seyferts.  Furthermore,
as Virgo Cluster galaxies were left out of the original sample, no data are
available for comparing field and cluster galaxies.  A larger 
near-infrared spectroscopic survey of a complete, volume-limited sample
of galaxies would provide additional constraints on these stochastic
bursts of star formation, such as the frequency of their occurence, their
relation to bars, environment, Seyfert activity, and LINER activity, and
the range of masses formed within these bursts.

Single slit spectrometers
are good for sampling the pointlike nuclei of many nearby galaxies and can
return spectroscopic data that accurately represent the stellar populations
and total star formation activity within these galaxies.  However, when the
nuclear star formation structures become more complex, as is the case for
NGC~3556 and NGC~4100, then a single slit could possibly
miss much of the spectral line emission coming from within the centers of
these galaxies.  Therefore, integral field near-infrared spectroscopic 
surveys of nearby spiral galaxies would appear to be warranted.  Such a survey 
would not only provide more accurate data on the stellar populations and 
total line emission from the nuclei of galaxies but also provide data on 
the number of galaxies with complex, extended nuclear structures
that could lead to additional clues on the phenomenology of nuclear star
formation activity.

\acknowledgements

GJB would like to thank Barry Rothberg, John Rayner, and Mark Seigar
for their assistance during observing; Michael Cusihing for his help
on data processing; Almudena Alonso-Herrero, Charles Engelbracht,
Elizabeth Barton Gillespie, Robert C.  Kennicutt, and Michael Meyer
for their helpful discussions on the analysis.  This research has been
supported by NASA grants NAG 5-3370 and JPL 961566.

\clearpage

\clearpage

\clearpage

\clearpage

\clearpage

\clearpage

\clearpage

\clearpage

\clearpage

\clearpage

\clearpage

\clearpage

\clearpage

\clearpage

\clearpage

\clearpage

\clearpage

\clearpage

\clearpage

\clearpage

\begin{deluxetable}{lccc}
\tablecolumns{4}
\tablewidth{0pc}
\tablecaption{Morphology, Distance, and Nuclear Activity Data for the Sample
\label{t_sample}}
\tablehead{
\colhead{Galaxy} &
\colhead{Morphological Type\tablenotemark{a}} &
\colhead{Distance (Mpc)\tablenotemark{b}} &
\colhead{Nuclear Activity\tablenotemark{c}}}
\startdata
NGC  289 & SB(rs)bc	&	19.4	&	\nodata \\
NGC 1512 & SB(r)a	&	9.5	&	\nodata \\
NGC 1569 & IBm		&	1.6	&	H       \\
NGC 3556 & SB(s)cd	&	14.1	&	H	\\
NGC 3898 & SA(s)ab	&	21.9	&	T2	\\
NGC 4088 & SAB(rs)bc	&	17.0	&	H	\\
NGC 4096 & SAB(rs)c	&	8.8	&	H	\\
NGC 4100 & PSA(rs)bc	&	17.0	&	H	\\
NGC 4157 & SAB(s)b	&	17.0	&	H	\\
NGC 4203 & SAB0		&	9.7	&	L1.9	\\
NGC 4244 & SA(s)cd	&	3.1	&	H	\\
NGC 4274 & RSB(r)ab	&	9.7	&	H	\\
NGC 4314 & SB(rs)a	&	9.7	&	L2	\\
NGC 4414 & SA(rs)c	&	9.7	&	T2:	\\
NGC 4448 & SB(r)ab	&	9.7	&	H	\\
NGC 4710 & SA(r)0	&	16.8	&	H	\\
NGC 4725 & SAB(r)ab pec	&	12.4	&	S2:	\\
NGC 4826 & RSA(rs)ab 	&	4.1	&	T2	\\
NGC 4984 & RSAB(rs)0 	&	21.3	&	\nodata	\\
NGC 5005 & SAB(rs)bc	&	21.3	&	L1.9	\\
NGC 5033 & SA(s)c	&	18.7	&	S1.5	\\
NGC 5054 & SA(s)bc	&	27.3	&	\nodata	\\
NGC 5055 & SA(rs)bc	&	7.2	&	T2	\\
NGC 5087 & SA0		&	27.8	&	\nodata	\\
NGC 5101 & RSB(rs)0/a	&	27.4	&	\nodata	\\
NGC 5102 & SA0		&	3.5	&	\nodata	\\
NGC 5170 & SA(s)c	&	24.0	&	\nodata	\\
NGC 5371 & SAB(rs)bc	&	37.8	&	L2	\\
NGC 5457 & SAB(rs)cd	&	5.4	&	H	\\
NGC 5566 & SB(r)ab	&	26.4	&	L2	\\
NGC 5676 & SA(rs)bc	&	34.5	&	H	\\
NGC 5701 & RSB(rs)0/a	&	26.1	&	T2:	\\
NGC 5713 & SAB(rs)bc pec&	30.4	&	\nodata	\\
NGC 5746 & SAB(rs)b	&	29.4	&	T2	\\
NGC 5792 & SB(rs)b	&	30.6	&	\nodata	\\
NGC 5838 & SA0		&	28.5	&	T2::	\\
NGC 5846 & E0		&	28.5	&	T2:	\\
NGC 5850 & SB(r)b	&	28.5	&	L2	\\
NGC 5866 & SA0		&	15.3	&	T2	\\
NGC 5907 & SA(s)c	&	14.9	&	H:	\\
NGC 5985 & SAB(r)b	&	39.2	&	L2	\\
\enddata
\tablenotetext{a}{As defined in RC3.}
\tablenotetext{b}{Designations as given in HFS97b.  S = Seyfert.
L = LINER.  H = HII nucleus.  T = Transitional.  The number indicates the type
of AGN activity.  Single colons indicate uncertain classifications.  Double
colons indicate highly uncertain classifications.  Note that the 
classifications are based strictly on optical line ratios and that not all
galaxies have been classified.}
\end{deluxetable}

\clearpage

\begin{deluxetable}{lcccc}
\tablecolumns{5}
\tablewidth{0pc}
\tablecaption{K-Band Observations Information \label{t_obsinfok}}
\tablehead{
\colhead{Target} &
\colhead{Date of} &
\colhead{Instrument} &
\colhead{Total Integration} &
\colhead{Spectroscopic}\\
\colhead{} &
\colhead{Observations} &
\colhead{} &
\colhead{Time (s)} &
\colhead{Standard Types}}
\startdata
NGC  289 &     19 Sep 2000 &     SPEX &     2400 &     A0V, G2Ib \\
NGC 1512 &     17 Sep 2000 &     SPEX &     1800 &     A0V       \\
NGC 1569 &     16 Sep 2000 &     SPEX &     2400 &     A0V, G2V  \\
NGC 3556 &     08 May 2001 &     CGS4 &     960  &     G2V       \\ 
NGC 3898 &     04 Apr 2001 &     SPEX &     1200 &     A0V, G5   \\
NGC 4088 &     08 May 2001 &     CGS4 &     960  &     G2V       \\
NGC 4096 &     10 May 2001 &     CGS4 &     960  &     G5V       \\
NGC 4100 &     09 May 2001 &     CGS4 &     960  &     G5V       \\
NGC 4157 &     09 May 2001 &     CGS4 &     960  &     G5V       \\
NGC 4203 &     07 May 2001 &     CGS4 &     960  &     G5V       \\
NGC 4244 &     08 May 2001 &     CGS4 &     960  &     G5V       \\
NGC 4274 &     07 May 2001 &     CGS4 &     960  &     G5V       \\
NGC 4314 &     07 May 2001 &     CGS4 &     960  &     G5V       \\
NGC 4414 &     07 Jun 2000 &     SPEX &     1200 &     A0V       \\
NGC 4448 &     07 May 2001 &     CGS4 &     960  &     G5V       \\
NGC 4710 &     10 May 2001 &     CGS4 &     960  &     G2V       \\   
NGC 4725 &     07 Jun 2000 &     SPEX &     1200 &     A0V       \\ 
NGC 4826 &     05 Jun 2000 &     SPEX &     2160 &     A0        \\ 
NGC 4984 &     10 May 2001 &     CGS4 &     960  &     G6V       \\
NGC 5005 &     05 Jun 2000 &     SPEX &     720  &     A0V       \\
NGC 5033 &     06 Jun 2000 &     SPEX &     1200 &     A0V       \\ 
NGC 5054 &     08 Jun 2000 &     SPEX &     1200 &     A0V       \\ 
NGC 5055 &     06 Jun 2000 &     SPEX &     1200 &     A0V       \\ 
NGC 5087 &     10 May 2001 &     CGS4 &     960  &     G6V       \\
NGC 5101 &     08 May 2001 &     CGS4 &     960  &     G2V       \\
NGC 5102 &     09 May 2001 &     CGS4 &     960  &     G3V       \\
NGC 5170 &     08 May 2001 &     CGS4 &     960  &     G6V       \\
NGC 5371 &     07 May 2001 &     CGS4 &     960  &     G5V       \\
NGC 5457 &     09 May 2001 &     CGS4 &     960  &     G5V       \\
NGC 5566 &     06 Jun 2000 &     SPEX &     720  &     A0V       \\ 
NGC 5676 &     07 May 2001 &     CGS4 &     960  &     G5V       \\
NGC 5701 &     08 May 2001 &     CGS4 &     960  &     G1V       \\
NGC 5713 &     07 May 2001 &     CGS4 &     960  &     G1V       \\
NGC 5746 &     06 Apr 2001 &     SPEX &     1560 &     A0V, G5   \\
NGC 5792 &     07 May 2001 &     CGS4 &     960  &     G1V       \\
NGC 5838 &     08 May 2001 &     CGS4 &     960  &     G1V       \\
NGC 5846 &     05 Apr 2001 &     SPEX &     1200 &     A0V, G5   \\
NGC 5850 &     08 May 2001 &     CGS4 &     960  &     G1V       \\
NGC 5866 &     07 May 2001 &     CGS4 &     960  &     G5V       \\
NGC 5907 &     08 May 2001 &     CGS4 &     960  &     G5V       \\
NGC 5985 &     09 May 2001 &     CGS4 &     960  &     G5V       \\
\enddata
\end{deluxetable}

\clearpage

\begin{deluxetable}{lcccc}
\tablecolumns{5}
\tablewidth{0pc}
\tablecaption{H-Band Observations Information \label{t_obsinfoh}}
\tablehead{
\colhead{Target} &
\colhead{Date of} &
\colhead{Instrument} &
\colhead{Total Integration} &
\colhead{Spectroscopic}\\
\colhead{} &
\colhead{Observations} &
\colhead{} &
\colhead{Time (s)} &
\colhead{Standard Types}}
\startdata
NGC  289 &     19 Sep 2000 &     SPEX &     2400 &     A0V, G2Ib \\
NGC 1512 &     17 Sep 2000 &     SPEX &     1800 &     A0V       \\ 
NGC 1569 &     16 Sep 2000 &     SPEX &     2400 &     A0V, G2V  \\ 
NGC 3898 &     04 Apr 2001 &     SPEX &     1200 &     A0V, G5   \\     
NGC 4414 &     07 Jun 2000 &     SPEX &     1200 &     A0V       \\ 
NGC 4725 &     07 Jun 2000 &     SPEX &     1200 &     A0V       \\ 
NGC 4826 &     05 Jun 2000 &     SPEX &     2160 &     A0        \\
NGC 5005 &     05 Jun 2000 &     SPEX &     720  &     A0V       \\ 
NGC 5033 &     06 Jun 2000 &     SPEX &     1200 &     A0V       \\ 
NGC 5054 &     08 Jun 2000 &     SPEX &     1200 &     A0V       \\ 
NGC 5055 &     06 Jun 2000 &     SPEX &     1200 &     A0V       \\  
NGC 5371 &     10 May 2001 &     CGS4 &     960  &     G5V       \\
NGC 5566 &     06 Jun 2000 &     SPEX &     720  &     A0V       \\ 
NGC 5676 &     10 May 2001 &     CGS4 &     960  &     G5V       \\
NGC 5713 &     10 May 2001 &     CGS4 &     960  &     G1V       \\
NGC 5746 &     06 Apr 2001 &     SPEX &     1560 &     A0V, G5   \\
NGC 5846 &     05 Apr 2001 &     SPEX &     1200 &     A0V, G5   \\
NGC 5866 &     10 May 2001 &     CGS4 &     1200 &     G5V       \\
NGC 5907 &     10 May 2001 &     CGS4 &     960  &     G5V       \\
NGC 5985 &     10 May 2001 &     CGS4 &     960  &     G5V       \\     
\enddata
\end{deluxetable}

\clearpage

\begin{deluxetable}{lcc}
\tablecolumns{3}
\tablewidth{0pc}
\tablecaption{Spectral Features Identified in the Quiescent Composite K-Band
  Spectrum \label{t_quieskline}}
\tablehead{
\colhead{Feature} & \colhead{Wavenumber}  & \colhead{Wavelength} \\
\colhead{}        & \colhead{(cm$^{-1}$)} & \colhead{($\mu$m)}}
\startdata
Mg {\small I} &              4747 &     2.107 \\
Mg {\small I} &              4747 &     2.107 \\
Al {\small I} &              4740 &     2.110 \\
Al {\small I} &              4724 &     2.117 \\
Na {\small I} &              4533 &     2.206 \\
Na {\small I} &              4527 &     2.209 \\
Fe {\small I} &              4492 &     2.226 \\
Fe {\small I} &              4467 &     2.239 \\
Ca {\small I} &              4422 &     2.261 \\
Ca {\small I} &              4419 &     2.263 \\
Ca {\small I} &              4414 &     2.266 \\
Mg {\small I} &              4386 &     2.281 \\
$^{12}$CO 2-0 bandhead &     4360 &     2.294 \\
$^{12}$CO 3-1 bandhead &     4305 &     2.323 \\
$^{13}$CO 2-0 bandhead &     4265 &     2.345 \\
$^{12}$CO 4-2 bandhead &     4251 &     2.352 \\
$^{13}$CO 3-1 bandhead &     4212 &     2.374 \\
$^{12}$CO 5-3 bandhead &     4196 &     2.383 \\
\enddata
\end{deluxetable}

\clearpage

\begin{deluxetable}{lcc}
\tablecolumns{3}
\tablewidth{0pc}
\tablecaption{Spectral Features Identified in the Quiescent Composite H-Band
  Spectrum \label{t_quieshline}}
\tablehead{
\colhead{Feature} & \colhead{Wavenumber}  & \colhead{Wavelength} \\
\colhead{}        & \colhead{(cm$^{-1}$)} & \colhead{($\mu$m)}}
\startdata
$^{12}$CO 3-0 bandhead &              6418 &     1.558 \\
$^{12}$CO 4-1 bandhead &              6337 &     1.578 \\
Si {\small I} &                       6292 &     1.589 \\
$^{12}$CO 5-2 bandhead &              6257 &     1.598 \\
$^{12}$CO 6-3 bandhead &              6177 &     1.619 \\
$^{12}$CO 7-4 bandhead &              6097 &     1.640 \\
$^{12}$CO 8-5 bandhead &              6018 &     1.662 \\
Al {\small I} &                       5977 &     1.673 \\
Al {\small I} &                       5968 &     1.676 \\
Al {\small I} &                       5964 &     1.677 \\
$^{12}$CO 9-6 bandhead &              5938 &     1.684 \\
$^{12}$CO 10-7 bandhead &             5959 &     1.707 \\
Mg {\small I} &                       5843 &     1.711 \\
$^{12}$CO 11-8 bandhead &             5780 &     1.730 \\
\enddata
\end{deluxetable}

\clearpage

\begin{deluxetable}{lcc}
\tablecolumns{3}
\tablewidth{0pc}
\tablecaption{Measurements of Absorption Line Features in the Composite 
  Quiescent Spectra \label{t_quiesmeas}}
\tablehead{
\colhead{Feature} & \colhead{Central Wavelength\tablenotemark{a}}  & 
     \colhead{Equivalent Width} \\
\colhead{}        & \colhead{($\mu$m)} & 
     \colhead{(\AA)}}
\startdata
Si {\small I}      &     1.5890 &    $2.37 \pm 0.05$\\ 
CO 6-3             &     1.6198 &    $3.55 \pm 0.04$\\
CO 2-0             &     2.2950 &    $9.95 \pm 0.03$\\
\enddata
\tablenotetext{a}{As defined by \citet{omo93}}
\end{deluxetable}

\clearpage

\begin{deluxetable}{lcccccccc}
\rotate
\tablecolumns{9}
\tablewidth{0pc}
\tabletypesize{\scriptsize}
\tablecaption{Measurement of Spectral Features in Non-Quiestcent Galaxies
\label{t_nonquiesmeas}}
\tablehead{
\colhead{Galaxy\tablenotemark{a}}& 
\multicolumn{2}{c}{Fe~II (1.6440~\micron)} &
\multicolumn{2}{c}{He~I (2.0587~\micron)} &
\multicolumn{2}{c}{H$_2$ 1-0 S(1) (2.1218~\micron)}&
\multicolumn{2}{c}{H~I Br-$\gamma$ (2.1661~\micron)}\\
\colhead{}&
\colhead{Eq. Width} &     \colhead{Flux} &
\colhead{Eq. Width} &     \colhead{Flux} &
\colhead{Eq. Width} &     \colhead{Flux} &
\colhead{Eq. Width} &     \colhead{Flux}\\
\colhead{}&
\colhead{(\AA)} &     \colhead{(W m$^{-2}$)} &
\colhead{(\AA)} &     \colhead{(W m$^{-2}$)} &
\colhead{(\AA)} &     \colhead{(W m$^{-2}$)} &
\colhead{(\AA)} &     \colhead{(W m$^{-2}$)}}
\startdata
NGC 289 &
     $2.2 \pm 0.3$ &	$1.6 \pm 0.2 \times10^{-19}$ &
     $0.5 \pm 0.2$ &	$2.1 \pm 0.8 \times10^{-20}$ &
     $3.8 \pm 0.3$ &	$1.6 \pm 0.1 \times10^{-19}$ &
     $4.1 \pm 0.3$ &	$1.5 \pm 0.1 \times10^{-19}$ \\
NGC 1569\tablenotemark{b} &
     $-0.4 \pm 0.2$ &	$-6.1 \pm 3.3 \times10^{-19}$ &
     $1.7 \pm 0.2$ &	$4.4 \pm 0.6 \times10^{-19}$ &
     $0.5 \pm 0.2$ &	$1.2 \pm 0.4 \times10^{-19}$ &
     $1.3 \pm 0.2$ &	$2.8 \pm 0.3 \times10^{-19}$ \\
NGC 3556\tablenotemark{c} &
     \nodata &		\nodata	&
     $15 \pm 2$ &	$6.8 \pm 0.8 \times10^{-19}$ &
     $4 \pm 2$ &	$1.7 \pm 0.7 \times10^{-19}$ &
     $28 \pm 2$ &	$1.20 \pm 0.07 \times10^{-18}$ \\
NGC 4088 &
     \nodata &		\nodata	&
     $-0.1 \pm 0.4$ &	$-0.2 \pm 1.2 \times10^{-19}$ &
     $0.4 \pm 0.4$ &	$1.1 \pm 1.2 \times10^{-19}$ &
     $2.8 \pm 0.4$ &	$7.2 \pm 1.1\times10^{-19}$ \\
NGC 5005 &
     $4.8 \pm 0.4$ &	$8.8 \pm 0.7 \times10^{-18}$ &
     $0.0 \pm 0.2$ &	$0.2 \pm 1.7 \times10^{-19}$ &
     $10.4 \pm 0.3$ &	$8.0 \pm 0.3 \times10^{-18}$ &
     $0.1 \pm 0.8$ &	$0.9 \pm 6.1 \times10^{-19}$ \\
NGC 5033 &
     $3.4 \pm 0.3$ &	$3.4 \pm 0.3 \times10^{-18}$ &
     $0.3 \pm 0.2$ &	$1.3 \pm 0.8 \times10^{-19}$ &
     $2.8 \pm 0.2$ &	$1.2 \pm 0.1 \times10^{-18}$ &
     $1.9 \pm 0.3$ &	$7.8 \pm 1.1 \times10^{-19}$ \\
NGC 5676\tablenotemark{d} &     
     \nodata &		\nodata &
     $23 \pm 6$ &	$4.0 \pm 1.1 \times10^{-19}$ &
     $5 \pm 8$  &       $0.8 \pm 1.4 \times10^{-19}$ &
     $70 \pm 9$ &	$1.1 \pm 0.1 \times10^{-18}$ \\
NGC 5713 &
     $7.1 \pm 0.9$ &	$1.7 \pm 0.2 \times10^{-18}$ &
     $1.3 \pm 0.9$ &	$1.9 \pm 1.3 \times10^{-19}$ &
     $2.8 \pm 0.7$ &	$3.8 \pm 0.9 \times10^{-19}$ &
     $5.9 \pm 1.3$ &	$7.5 \pm 1.7 \times10^{-19}$ \\
\enddata
\tablenotetext{a}{The nuclear region is used unless otherwise noted.}
\tablenotetext{b}{For cluster A.}
\tablenotetext{c}{For the west nuclear region.}
\tablenotetext{d}{For the northern knot.}
\end{deluxetable}

\clearpage

\begin{deluxetable}{lcc}
\tablecolumns{3}
\tablewidth{0pc}
\tablecaption{Parameters Describing the Underlying Power Law Emission in 
NGC~5005 and NGC~5033 \label{t_n5005-5033pw}}
\tablehead{
\colhead{Galaxy} &   \colhead{Spectral Index} &   \colhead{Percent of} \\
\colhead{} &         \colhead{($\alpha$ in $\lambda^{-\alpha}$)} &
\colhead{K-band Flux}}
\startdata
NGC~5005 &     1.7 &    7\% \\
NGC~5033 &     1.7 &    22\% \\
\enddata
\end{deluxetable}

\clearpage

\begin{deluxetable}{ccccc}
\tablecolumns{5}
\tablewidth{0pc}
\tablecaption{Brackett-$\gamma$ Luminosities and Luminosity-to-Mass Conversion
Factors \label{t_brgconvert}}
\tablehead{
\colhead{Time} &
\colhead{Luminosity} &       
\colhead{Median} &
\colhead{Median Mass} &  
\colhead{Luminosity-to-Mass} \\
\colhead{Range} &
\colhead{Range\tablenotemark{a}} &
\colhead{Luminosity\tablenotemark{a}} &
\colhead{Remaining\tablenotemark{a}} &
\colhead{Conversion} \\
\colhead{(Myr)} &
\colhead{ \ (W)} &                    
\colhead{(W)} &
\colhead{(M$_\odot$)} &
\colhead{Factor (M$_\odot$ W$^{-1}$)}}
\startdata
0.0 - 3.5 &    
  $3.1 \times 10^{31}$ - $9.0 \times 10^{31}$ &     $8.6 \times 10^{31}$ &
  $9.9 \times 10^{5}$ &                             $1.2 \times 10^{-26}$ \\
3.5 - 8.0 &
  $6.5 \times 10^{29}$ - $3.0 \times 10^{31}$ &     $3.9 \times 10^{30}$ &
  $9.4 \times 10^{5}$ &                             $2.4 \times 10^{-25}$ \\
\enddata
\tablenotetext{a}{For a burst that produces $10^6$~M$_\odot$ of stars.}
\end{deluxetable}

\clearpage

\begin{deluxetable}{cccccc}
\tablecolumns{6}
\tablewidth{0pc}
\tablecaption{Fe~{\small II} 1.644~$\mu$m Line Luminosities and 
Luminosity-to-Mass Conversion Factors \label{t_feiiconvert}}
\tablehead{
\colhead{Time} &
\colhead{Supernova} &   
\colhead{Median} &   
\colhead{Median} &
\colhead{Median} &  
\colhead{Luminosity-to-Mass} \\
\colhead{Range} &
\colhead{Rate} &
\colhead{Supernova} &
\colhead{Luminosity\tablenotemark{a}} &
\colhead{Mass} &
\colhead{Conversion} \\
\colhead{(Myr)} &
\colhead{Range\tablenotemark{a}} &
\colhead{Rate\tablenotemark{a}} &
\colhead{(W)} &
\colhead{Remaining\tablenotemark{a}} &
\colhead{Factor} \\
\colhead{} &
\colhead{(yr$^{-1}$)} &            
\colhead{(yr$^{-1}$)} &
\colhead{} &
\colhead{(M$_\odot$)} &
\colhead{(M$_\odot$ W$^{-1}$)}}
\startdata
3.5 - 8.0 &
  $7.7 \times 10^{-4}$ - $1.8 \times 10^{-3}$ &     $8.8 \times 10^{-4}$ &
  $1.1 \times 10^{31}$ &
  $9.4 \times 10^{5}$ &                             $8.5 \times 10^{-26}$ \\
8.0 - 36 &
  $3.5 \times 10^{-4}$ - $7.6 \times 10^{-4}$ &     $4.6 \times 10^{-4}$ &
  $5.5 \times 10^{30}$ &
  $8.3 \times 10^{5}$ &                             $1.5 \times 10^{-25}$ \\
\enddata
\tablenotetext{a}{For a burst that produces $10^6$~M$_\odot$ of stars.}
\end{deluxetable}

\clearpage

\begin{deluxetable}{ccccc}
\tablecolumns{5}
\tablewidth{0pc}
\tablecaption{Ratios of K-band Luminosities to Line Luminosities
\label{t_contlinerat}}
\tablehead{
\colhead{Time Range} &
\colhead{K-Band} &
\colhead{Median K-Band} &
\colhead{K-Band} &
\colhead{K-Band} \\
\colhead{(Myr)} &
\colhead{Luminosity} &
\colhead{Luminosity\tablenotemark{a} } &
\colhead{/ Brackett-$\gamma$} &
\colhead{/ Fe~{\small II}} \\
\colhead{} &
\colhead{Range\tablenotemark{a} \ (W)} &
\colhead{(W)} &
\colhead{Ratio} &
\colhead{Ratio}}
\startdata
0.0 - 3.5 & 
     $3.7 \times 10^{32}$ - $6.1 \times 10^{32}$ &     $5.8 \times 10^{32}$ &
     6.7 &        \nodata \\
3.5 - 8.0 &
     $2.9 \times 10^{32}$ - $1.7 \times 10^{33}$ &     $4.5 \times 10^{32}$ &
     120 &        41 \\
8.0 - 36 &
     $2.4 \times 10^{32}$ - $2.3 \times 10^{33}$ &     $3.1 \times 10^{32}$ &
     \nodata &    56 \\
\enddata
\tablenotetext{a}{For a burst that produces $10^6$~M$_\odot$ of stars.}
\end{deluxetable}

\clearpage

\begin{deluxetable}{lcccccccccc}
\rotate
\tablecolumns{10}
\tablewidth{0pc}
\tabletypesize{\scriptsize}
\tablecaption{Young and Old Stellar Masses and Their Ratios for Non-Quiescent
Galaxies \label{t_youngoldrat}}
\tablehead{
\colhead{Galaxy\tablenotemark{a}} &
\colhead{Age} &
\multicolumn{2}{c}{Spectral Line} &
\colhead{Young} &
\multicolumn{4}{c}{K Continuum} &
\colhead{Old} &
\colhead{Young} \\
\colhead{} &
\colhead{Range} &
\colhead{} &                         \colhead{} &
\colhead{Stellar} &
\colhead{Total} &                    \colhead{From} &     
     \colhead{From} &                \colhead{From} &
\colhead{Stellar} &
\colhead{/ Old} \\
\colhead{} &
\colhead{} &
\colhead{Name} &                     \colhead{Luminosity} &
\colhead{Mass} &
\colhead{System} &                   \colhead{Young} &     
     \colhead{Power-Law} &           \colhead{Old} &
\colhead{Mass} &
\colhead{Mass} \\
\colhead{} &
\colhead{(Myr)} &
\colhead{}&                          \colhead{(W)} &
\colhead{(M$_\odot$)} &
\colhead{(W)} &                      \colhead{Stars (W)} &      
     \colhead{Continuum (W)} &       \colhead{Stars (W)} &
\colhead{(M$_\odot$)} &
\colhead{Ratio}

}
\startdata
NGC~289 &
     3.5 - 8.0 &
     Brackett-$\gamma$ &       $6.8 \times 10^{29}$ &
     $1.6 \times 10^5$ &
     $6.8 \times 10^{32}$ &    $8.2 \times 10^{31}$ &
     \nodata &                 $6.0 \times 10^{32}$ &
     $7.8 \times 10^6$ &
     $2.1 \times 10^{-2}$ \\
&
     &
     Fe~II &                   $7.2 \times 10^{29}$ &
     $6.1 \times 10^4$ &
     $6.8 \times 10^{32}$ &    $3.0 \times 10^{31}$ &
     \nodata &                 $6.5 \times 10^{32}$ &
     $8.5 \times 10^6$ &
     $7.2 \times 10^{-3}$ \\
NGC~1569\tablenotemark{b} &
     0.0 - 3.5 &
     Brackett-$\gamma$ &       $8.6 \times 10^{27}$ &
     $1.0 \times 10^2$ &
     $2.7 \times 10^{31}$ &    $5.8 \times 10^{28}$ &
     \nodata &                 $2.7 \times 10^{31}$ &
     $3.5 \times 10^5$ &
     $2.9 \times 10^{-4}$ \\
NGC~3556\tablenotemark{c} &
     0.0 - 3.5 &
     Brackett-$\gamma$ &       $2.9 \times 10^{30}$ &
     $3.5 \times 10^4$ &
     $4.0 \times 10^{32}$ &    $1.9 \times 10^{31}$ &
     \nodata &                 $3.8 \times 10^{32}$ &
     $4.9 \times 10^6$ &
     $7.1 \times 10^{-3}$ \\
NGC~4088 &
     0.0 - 3.5 &
     Brackett-$\gamma$ &       $2.5 \times 10^{30}$ &
     $3.0 \times 10^4$ &
     $3.5 \times 10^{33}$ &    $1.7 \times 10^{31}$ &
     \nodata &                 $3.5 \times 10^{33}$ &
     $4.6 \times 10^7$ &
     $6.5 \times 10^{-4}$ \\
NGC~5005 &
     8.0 - 36 &
     Fe~II &                   $4.8 \times 10^{31}$ &
     $7.2 \times 10^6$ &
     $1.6 \times 10^{34}$ &    $2.7 \times 10^{33}$ &
     $1.1 \times 10^{33}$ &    $1.2 \times 10^{34}$ &
     $1.6 \times 10^8$ &
     $4.5 \times 10^{-2}$ \\
NGC~5033 &
     8.0 - 36 &
     Fe~II &                   $1.4 \times 10^{31}$ &
     $2.1 \times 10^6$ &
     $6.7 \times 10^{33}$ &    $7.8 \times 10^{32}$ &
     $1.5 \times 10^{33}$ &    $4.4 \times 10^{33}$ &
     $5.7 \times 10^7$ &
     $3.7 \times 10^{-2}$ \\
NGC~5676\tablenotemark{d} &
     0.0 - 3.5 &
     Brackett-$\gamma$ &       $1.6 \times 10^{31}$ &
     $1.9 \times 10^5$ &
     $9.5 \times 10^{32}$ &    $1.1 \times 10^{32}$ &
     \nodata &                 $8.4 \times 10^{32}$ &
     $1.1 \times 10^7$ &
     $1.7 \times 10^{-2}$ \\
NGC~5713 &
     3.5 - 8.0 &
     Brackett-$\gamma$ &       $8.3 \times 10^{30}$ &
     $2.0 \times 10^6$ &
     $5.6 \times 10^{33}$ &    $1.0 \times 10^{33}$ &
     \nodata &                 $4.6 \times 10^{33}$ &
     $6.0 \times 10^7$ &
     $3.3 \times 10^{-2}$ \\
&
     &
     Fe~II &                   $1.9 \times 10^{31}$ &
     $1.6 \times 10^6$ &
     $5.6 \times 10^{33}$ &    $7.8 \times 10^{32}$ &
     \nodata &                 $4.8 \times 10^{33}$ &
     $6.2 \times 10^7$ &
     $2.6 \times 10^{-2}$ \\
\enddata
\tablenotetext{a}{The nuclear region is used unless otherwise noted.}
\tablenotetext{b}{For cluster A.}
\tablenotetext{c}{For the west nuclear region.}
\tablenotetext{d}{For the northern knot.}
\end{deluxetable}


\clearpage

\begin{thebibliography}{}
\bibitem[Alonso-Herrero et al.(1997)]{aetal97}Alonso-Herrero, A., Rieke, M. J.,
        Rieke, G. H., \& Ruiz, M.  1997, \apj, 482, 747
\bibitem[Alonso-Herrero et al.(2000)]{aetal00}Alonso-Herrero, A., Rieke, M. J.,
        Rieke, G. H., \& Shields, J. C.  2000, \apj, 530, 688
\bibitem[Alonso-Herrero et al.(2003)]{aetal03}Alonso-Herrero, A., Rieke, G. H.,
        Rieke, M. J., \& Kelly, D. M.  2003, \aj, 125, 1210
\bibitem[Aloisi et al.(2001)]{aetal01}Aloisi, A., et al.  2001, \aj, 121, 1425
\bibitem[Arp (1981)]{a81}Arp, H.  1981, \apjs, 46, 75
\bibitem[Athanassoula (1992)]{a92}Athanassoula, E.  1992, \mnras, 259, 345
\bibitem[Bendo et al.(2002a)]{betal02a}Bendo, G. J., et al.  2002a, \aj, 
        123, 3067 (Paper 1)
\bibitem[Bendo et al.(2002b)]{betal02b}Bendo, G. J., et al.  2002b, \aj, 
        124, 1380 (Paper 2)
\bibitem[Boisson et al.(2002)]{betal02}Boisson, C., Coup\`{e}, S., Cuby, J. G.,
        Joly, M., \& Ward, M. J.  2002, \aap, 396, 489
\bibitem[Burston, Ward, \& Davies (2001)]{bwd01}Burston, A. J., Ward, M. J.,
        \& Davies, R. I.  2001, \mnras, 326, 403
\bibitem[Colina (1993)]{c93}Colina, L. 1993, \apj, 411, 565
\bibitem[Coziol, Doyon, \& Demers (2001)]{cdd01}Coziol, R., Doyon, R., \& 
        Demers, S.  2001, \mnras, 325, 1081
\bibitem[Cutri \& McAlary (1985)]{cm85}Cutri, R. M., \& McAlary, C. W.  1985,
        \apj, 296, 90
\bibitem[Dallier, Boisson, \& Joly (1996)]{dbj96}Dallier, R., Boisson, C., \&
        Joly, M.  1996, \aaps, 116, 239
\bibitem[Davies, Davidson, \& Johnson (1980)]{ddj80}Davies, R. D.,
	Davidson, G. P., \& Johnson, S. C.  1980, \mnras, 191, 253
\bibitem[de Vaucouleurs et al.(1991)]{detal91}de Vaucouleurs, G.,
	de Vaucouleurs, A., Corwin, H. G., Buta, R. J., Paturel, G., \&
	Fouque, P.  1991, Third Reference Catalogue of Bright Galaxies
	(Berlin: Springer-Verlag) (RC3)
\bibitem[Devereux (1987)]{d87}Devereux, N.  1987, \apj, 323, 91
\bibitem[Dultzin-Hacyan, Moles, \& Masegosa (1988)]{dmm88}Dultzin-Hacyan, D.,
	Moles, M., \& Masegosa, J.  1988, \aap, 206, 95
\bibitem[Engelbracht (1997)]{e97}Engelbracht, C. W.  1997, Ph.D. thesis, 
        University of Arizona
\bibitem[Forbes \& Ward (1993)]{fw93}Forbes, D. A., \& Ward, M. J. 1993, \apj,
        416, 150
\bibitem[F\"orster Schreiber (2000)]{f00}F\"orster Schreiber, N. M.  2000,
        \aj, 120, 2089
\bibitem[Friedli \& Benz (1993)]{fb93}Friedli, D., \& Benz, W.  1993, \aap,
        268, 65
\bibitem[Goldader et al.(1995)]{getal95}Goldader, J. D., Joseph, R. D., 
	Doyon, R., \& Sanders, D. B.  1995, \apj, 444, 97
\bibitem[Goldader et al.(1997)]{gjdetal97}Goldader, J. D., Joseph, R. D., 
	Doyon, R., \& Sanders, D. B.  1997, \apjs, 108, 449
\bibitem[Gonzalez-Delgado \& Perez (1993)]{gp93}Gonzalez-Delgado, R. M., \&
	Perez, E.  1993, \apss, 205, 127
\bibitem[Gonz\'{a}lez Delgado et al.(1998)]{ghletal98}Gonz\'{a}lez Delgado,
	R. M., Heckman, T., Leitherer, C., Meurer, G., Krolik, J.,
	Wilson, A. S., Kinney, A., \& Koratkar, A.  1998, \apj, 505, 174
\bibitem[Gonz\'{a}lez Delgado, Heckman, \& Leitherer (2001)]{ghl01}
        Gonz\'{a}lez Delgado, R. M., Heckman, T., \& Leitherer, C.  
        2001, \apj, 546, 845
\bibitem[Goorvitch (1994)]{g94}Goorvitch, D.  1994, \apjs, 95, 535
\bibitem[Greenhouse et al.(1991)]{getal91}Greenhouse, M. A., Woodward, C. E.,
        Thronson, H. A., Rudy, R. J., Rossano, G. S., Erwin, P., \& Puetter,
        R. C.  1991, \apj, 383, 164
\bibitem[Greenhouse et al.(1997)]{getal97}Greenhouse, M. A., et al. 1997,
        \apj, 476, 105
\bibitem[Greggio et al.(1998)]{gtcetal98}Greggio, L., Tosi, M., Clampin, M.,
        De Marchi, G., Leitherer, C., Nota, A., \& Sirianni, M.  1998, \apj,
        504, 725
\bibitem[Greve et al.(2002)]{getal02}Greve, A., Tarchi, A., H\"uttemeister, S.,
        de Grijs, R., van der Hulst, J. M., Garrington, S. T., Neininger, N.
        2002, \aap, 381, 825
\bibitem[Hawarden et al.(1979)]{hetal79}Hawarden, T. G., van Woerden, H.,
	Mebold, U., Goss, W. M., \& Peterson, B. A.  1979, \aap, 76, 230
\bibitem[Haynes (1979)]{h79}Haynes, M. P.  1979, \aj, 84, 1830
\bibitem[Heckman (1980)]{h80}Heckman, T. M.  1980, \aap, 87, 152
\bibitem[Heckman et al.(1989)]{hetal89}Heckman, T. M., Blitz, L., Wilson,
	A. S., Armus, L., \& Miley, G. K.  1989, \apj, 342, 735
\bibitem[Heckman et al.(1997)]{hetal97}Heckman, T. M., Gonzalez-Delgado, R., 
        Leitherer, C. Meurer, G. R., Krolik, J., Wilson, A. S., Koratkar, A., 
        \& Kinney, A.  1997, \apj, 482, 114
\bibitem[Heller \& Shlosman (1994)]{hs94}Heller, C. H., \& Shlosman, I.  1994,
        \apj, 424, 84
\bibitem[Helou, Salpeter, \& Terzian (1982)]{hst82}Helou, G. H.,
	Salpeter, E. E., \& Terzian, Y.  1982, \aj, 87, 1443
\bibitem[Hernquist \& Mihos (1995)]{hm95}Hernquist, L., \& Mihos, C. J.  1995,
        \apj, 448, 41
\bibitem[Hill et al.(1999)]{hetal99}Hill, T. L., Heisler, C. A., Sutherland,
        R., \& Hunstead, R. W.  1999, \aj, 117, 111
\bibitem[Ho, Filippenko, \& Sargent (1993)]{hfs93}Ho, L. C.,
	Filippenko, A. V., \& Sargent, W. L. W.  1993, \apj, 417, 63
\bibitem[Ho, Filippenko, \& Sargent (1997a)]{hfs97a}Ho, L. C., Filippenko,
	A. V., \& Sargent, W. L. W.  1997a, \apj, 487, 591
\bibitem[Ho, Filippenko, \& Sargent (1997b)]{hfs97b}Ho, L. C., Filippenko,
	A. V., \& Sargent, W. L. W.  1997b, \apjs, 112, 315 (HFS97b)
\bibitem[Huang et al.(1996)]{hetal96}Huang, J. H., Gu, Q. S., Su, H. J.,
	Hawarden, T. G., Liao, X. H., \& Wu, G. X.  1996, \aap, 313, 13
\bibitem[Ivanov (2000)]{i00}Ivanov, V. D.  2000, Ph.D. thesis, University of
        Arizona
\bibitem[Keel et al.(1985)]{ketal85}Keel, W. C., Kennicutt, R. C., Hummel, E.,
        \& van der Hulst, J. M.  1985, \aj, 90, 708
\bibitem[Kennicutt (1998)]{k_rc98}Kennicutt, R. C.  1998, \araa, 36, 189
\bibitem[Kennicutt et al.(1987)]{ketal87}Kennicutt, R. C., Roettiger, K. A.,
        Keel, W. C., van der Hulst, J. M., \& Hummel, E.  1987, \aj, 93, 1011
\bibitem[Kleinmann \& Hall (1986)]{kh86}Kleinmann, S. G., \& Hall, D. N. B.  
        1986, \apjs, 62, 501
\bibitem[Kudritzki (1998)]{k_rp98}Kudritzki, R. P. 1998, in Stellar 
        Astrophysics for the Local Group, ed. A. Aparicio, A. Herrero, \& F. 
        Sanchez (Cambridge: Cambridge Univ. Press), 149
\bibitem[Lan\c{c}on \& Rocca-Volmerange (1992)]{lr92}Lan\c{c}on, A., 
        \& Rocca-Volmerange, B.  1992, \aaps, 96, 593
\bibitem[Larkin et al.(1998)]{letal98}Larkin, J. E., Armus, L., Knop, R. A.,
	Soifer, B. T., \& Matthews, K.  1998, \apjs, 114, 59
\bibitem[Leitherer et al.(1999)]{letal99}Leitherer, Claus, et al.  1999, \apjs,
        123, 3 (Starburst99)
\bibitem[Mannucci et al.(2001)]{mbpetal01}Mannucci, F., Basile, F., 
        Poggianti, B. M., Cimatti, A., Daddi, E., Pozzetti, L., \& Vanzl, L.
        2001, \mnras, 326, 745 
\bibitem[Maoz, Ho, \& Sternberg (2001)]{mhs01}Maoz, D., Ho, L. C., \& 
        Sternberg, A.  2001, \apj, 554, L139
\bibitem[Meyer et al.(1998)]{metal98}Meyer, M. R., Edwards, S., Hinkle, K. H., 
        \& Strom, S. E.  1998, \apj, 508, 397
\bibitem[Mihos \& Hernquist (1994)]{mh94}Mihos, J. C., \& Hernquist, L.
        1994, \apj, 425, L13
\bibitem[Mihos \& Hernquist (1996)]{mh96}Mihos, J. C., \& Hernquist, L.
        1996, \apj, 464, 641
\bibitem[Morel, Doyon, \& St-Louis (2002)]{mds02}Morel, T., Doyon, R., \&
        St-Louis, N.  2002, \mnras, 329, 398
\bibitem[Mouri (1994)]{m94}Mouri, H.  1994, \apj, 427, 777
\bibitem[Murphy et al.(1999)]{metal99}Murphy, T. W., Soifer, B. T., Matthews,
        K., Kiger, J. R., \& Armus, L.  1999, \apj, 525, 85
\bibitem[Murphy et al.(2001)]{msmetal01}Murphy, T. W., Soifer, B. T., Matthews,
        K., Armus, L., \& Kiger, J. R.  2001, \aj, 121, 97
\bibitem[Noguchi (1988)]{n88}Noguchi, M.  1988, \aap, 203, 259
\bibitem[Oliva, Moorwood, \& Danziger (1989)]{omd89}Oliva, E., Moorwood,
        A. F. M., \& Danziger, I. J.  1989, \aap, 214, 307
\bibitem[Oliva et al.(1995)]{oetal95}Oliva, E., Origlia, L., Maiolino, R., \&
        Moorwood, A. F. M.  1995, \aap, 301, 55
\bibitem[Oliva et al.(1999)]{oetal99}Oliva, E., Origlia, L., Maiolino, R., \&
        Moorwood, A. F. M.  1999, \aap, 350, 9
\bibitem[Origlia, Moorwood, \& Oliva (1993)]{omo93}Origlia, L., 
        Moorwood, A. F. M., \& Oliva, E.  1993, \aap, 280, 536
\bibitem[Origlia \& Oliva (2000)]{oo00}Origlia, L., \& Oliva, E.  2000, \aap,
        357, 61
\bibitem[Origlia et al.(2001)]{oetal01}Origlia, L., Leitherer, C., Aloisi, A.,
        Greggio, L., \& Tosi, M.  2001, \aj, 122, 815
\bibitem[Rayner et al.(2003)]{retal03}Rayner, J. T., Toomey, D. W., 
        Onaka, P. M., Denault, A. J., Stahlberger, W. E., Vacca, W. D.,  
        Cushing, M. C., \&  Wang S. 2003, \pasp, in press
\bibitem[Reunanen, Kotilainen, \& Prieto (2002)]{rkp02}Reunanen, J.,
        Kotilainen, J. K., \& Prieto, M. A.  2002, \mnras, 331, 154 
\bibitem[Rodriguez Espinosa, Rudy, \& Jones (1987)]{rrj87}Rodriguez Espinosa, 
        J. M., Rudy, R. J., \& Jones, B.  1987, \apj, 312, 555
\bibitem[Roussel et al.(2001)]{retal01}Roussel, H., et al.  2001, \aap, 369,
	473
\bibitem[Rodnick, Rix, \& Kennicutt (2000)]{rrk00}Rudnick, G., Rix, H.-W., \&
        Kennicutt, R. C.  2000, \apj, 538, 569
\bibitem[Sandage, \& Bedke (1994)]{sb94}Sandage, A., \& Bedke, J.  1994,
	Canegie Atlas of Galaxies (Washington: Carnegie Institution)
\bibitem[Sandage \& Tammann (1987)]{st87}Sandage, A., \& Tammann, G. A.  1987,
	A Revised Shapley-Ames Catalog of Bright Galaxies 2nd ed. (Washington:
	Carnegie Institution)
\bibitem[Shields (1992)]{s92}Shields, J. C.  1992, \apj, 399, L27
\bibitem[Sosa-Brito, Tacconi-Garman, \& Lehnert (2001)]{stl01}Sosa-Brito, 
        R. M., Tacconi-Garman, L. E., \& Lehnert, M. D.  2001, \apjs, 136, 61
\bibitem[Terlevich \& Melnick (1985)]{tm85}Terlevich, R., \& Melnick, J.
        1985, \mnras, 213, 841
\bibitem[van der Werf et al.(1993)]{vetal93}van der Werf, P. P., Genzel, R.,
        Krabbe, A., Blietz, M., Lutz, D., Drapatz, S., Ward, M. J., \&
        Forb, D. A.  1993, \apj, 405, 522
\bibitem[van Moorsel (1983)]{v83}van Moorsel, G. A.  1983, \aaps, 54, 19
\bibitem[Vanzi, Alonso-Herrero, \& Rieke (1998)]{var98}Vanzi, L., 
        Alonso-Herrero, A., \& Rieke, G. H.  1998, \apj, 504, 93
\bibitem[Vanzi \& Rieke (1997)]{vr97}Vanzi, L., \& Rieke, G. H. 1997, \apj,
        479, 694
\bibitem[Wada \& Habe (1992)]{wh92}Wada, K., \& Habe, A.  1992, \mnras, 258,
        82
\bibitem[Wada \& Habe (1995)]{wh95}Wada, K., \& Habe, A.  1995, \mnras, 277,
        433
\bibitem[Wallace \& Hinkle (1997)]{wh97}Wallace, L., \& Hinkle, K.  1997,
        \apjs, 111, 445
\bibitem[Wallace \& Livingston (1992)]{wl92}Wallace, L., \& Livingston, W.
        1992, An Atlas of a Dark Sunspot Umbral Spectrum from 1970 to 
        8640~cm$^{-1}$ (1.16 to 5.1~$\mu$m), NSO Technical Report \#92-001 
	(Tucson: National Solar Observatory)
\bibitem[Wright et al.(1988)]{wetal88}Wright, G. S., Joseph, R. D., Robertson,
        N. A., James, P. A., \& Meikle, W. P. S.  1988, \mnras, 233, 1
\end{thebibliography}
\end{document}